\documentclass[aps,prx,twocolumn,longbibliography,amssymb,superscriptaddress,floatfix]{revtex4-1}
\usepackage{graphicx}
\usepackage{bm}
\usepackage{amsmath}
\usepackage{amssymb}
\usepackage{mathtools}
\usepackage{amsfonts}
\usepackage{cancel}
\usepackage{bbm}
\usepackage{mathbbol}
\usepackage{lipsum}
\usepackage{xspace}

\usepackage[
colorlinks=true,
urlcolor=blue,
linkcolor=blue,
citecolor=blue
]{hyperref}

\newcommand{\sys}{(a)}
\newcommand{\triad}{(b)}
\newcommand{\joseph}{(c)}

\newcommand{\avg}[1]{{\left\langle #1\right\rangle}}
\newcommand{\tavg}[1]{{\langle #1\rangle}}
\newcommand{\dd}{\mathbf{d}}
\newcommand{\vm}{\mathbf{m}}
\newcommand{\B}{\mathbf{B}}
\newcommand{\vsigma}{\bm{\sigma}}
\newcommand{\A}{\mathbf{A}}
\newcommand{\E}{\mathbf{E}}
\newcommand{\T}{\mathcal{E}}
\newcommand{\J}{\mathbf{J}}
\newcommand{\rr}{\mathbf{r}}
\newcommand{\Tr}[1]{{\mathrm{Tr}\left[#1\right]}}
\newcommand{\defeq}{\overset{\scriptscriptstyle\mathrm{def}}{=}}
\newcommand{\Helium}{\textsuperscript{3}He\xspace}
\renewcommand{\vec}[1]{{\bf #1}}

\newcommand{\mytitle}{Spin-polarized superconductivity: order parameter topology, current dissipation, and multiple-period
Josephson effect}

\begin{document}
    
\title{\mytitle}
    
\author{Eyal Cornfeld}
\affiliation{Department of Condensed Matter Physics, Weizmann Institute of Science, Rehovot 7610001, Israel}

\author{Mark S. Rudner}
\affiliation{Niels Bohr International Academy and Center for Quantum Devices, University of Copenhagen, 2100 Copenhagen, Denmark}
    
\author{Erez Berg}
\affiliation{Department of Condensed Matter Physics, Weizmann Institute of Science, Rehovot 7610001, Israel}

\begin{abstract}
We discuss transport properties of fully spin-polarized triplet superconductors, where only electrons of one spin component (along a certain axis) are paired. Due to the structure of the order parameter space, wherein phase and spin rotations are intertwined, a configuration where the superconducting phase winds by $4\pi$ in space is topologically equivalent to a configuration with no phase winding. This opens the possibility of supercurrent relaxation by a smooth deformation of the order parameter, where the order parameter remains non-zero at any point in space throughout the entire process. During the process, a spin texture is formed. We discuss the conditions for such processes to occur and their physical consequences. In particular, we show that when a voltage is applied, they lead to an unusual alternating-current Josephson effect whose period is an integer multiple of the usual Josephson period. These conclusions are substantiated in a simple time-dependent Ginzburg-Landau model for the dynamics of the order parameter. One of the potential applications of our analysis is for 
moir\'{e} systems, such as twisted bilayer and double bilayer graphene, where superconductivity is found in the vicinity of ferromagnetism.  
\end{abstract}

\maketitle

\section{Introduction}\label{sec:intro}
Spin-triplet superconductors (SCs) and superfluids are predicted to exhibit rich phenomena owing to the interplay between the spin and phase degrees of freedom of their order parameters. A celebrated example is superfluidity in \Helium~\cite{vollhardt2013superfluid,volovik2003universe}. 
Triplet superconductivity remains scarce in electronic systems, however;
possible examples include uranium heavy-fermion compounds where superconductivity is found to coexist with ferromagnetism~\cite{saxena2000superconductivity,aoki2001coexistence,Huy2007}, and Sr$_2$RuO$_4$~\cite{Mackenzie2003}, although the latter has recently been contested~\cite{pustogow2019constraints}. 

Two-dimensional moir\'{e} materials, such as twisted bilayer graphene (TBG) and heterostructures based on other van der Waals materials have recently emerged as a fertile ground for novel correlated-electron phenomena~\cite{DosSantos2007,bistritzer2011moire,li2010observation,cao2018correlated,cao2018unconventional,Kerelsky2019,Yankowitz1059,lu2019superconductors,Jiang2019,xie2019spectroscopic,choi2019electronic,Zondiner2019,Wong2020cascade,Stepanov2019,Sharpe2019,chen2019evidence,chen2019signatures,uri2020mapping,Cao2020strange,Cao2020,regan2020mott,chen2020tunable,regan2020mott}. In particular, the phase diagrams of these systems include spin and valley polarized states~\cite{Liu_2019,Sharpe2019,Zondiner2019} residing in proximity to superconducting states, raising the possibility of spin-triplet superconductivity. Moreover, since these systems have multiple valleys in their band structures and can be made relatively clean, they may avoid the pair-breaking effect of disorder that inhibits triplet superconductivity in many materials. In carbon-based materials, spin-orbit coupling is expected to be negligible, opening the possibility of a non-trivial intertwining of the gapless magnetic and superconducting phase degrees of freedom. Intriguingly, recent experiments in twisted double-bilayer graphene (TDBG) with a perpendicular electric field show possible signs of triplet superconductivity~\cite{liu2019spin,shen2020correlated}.

%%%%%%%%%%%%%%%%%%%%%%%%%%%%%%%
\begin{figure}[t!]
\begin{center}
\includegraphics[width=1\linewidth]{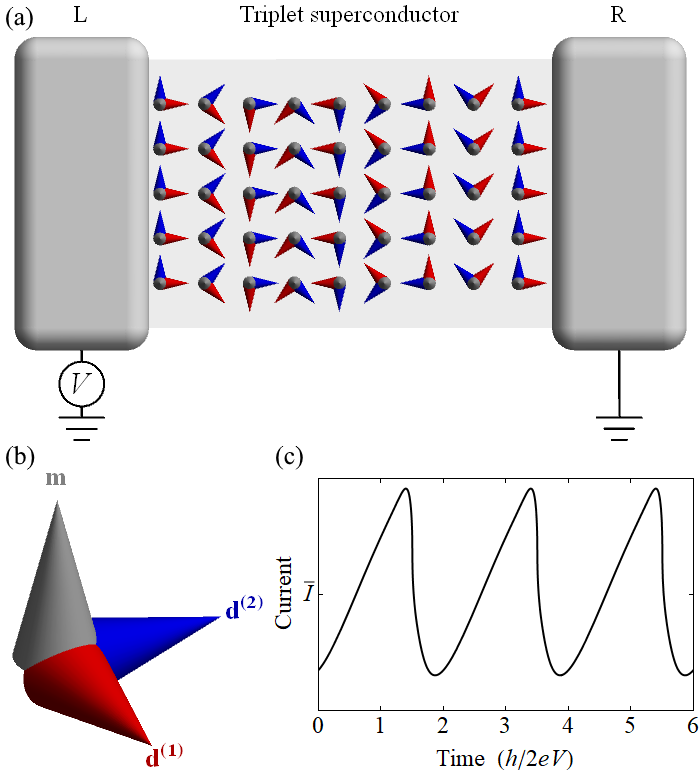}
\end{center}
\caption{
(a) A depiction of the proposed experimental setup; a fully spin-polarized triplet superconductor is connected to a DC voltage source and drain via the left and right leads. We overlay a current-carrying configuration of the order parameter.
(b) A visualization of the order parameter triad where $\dd^{(1)}$ and $\dd^{(2)}$ are depicted by the red and blue arrows and the pairing polarization $\vm$ is depicted by the gray arrow.
(c) Schematics of the current response of the system portray AC Josephson effect period doubling.
}
\label{fig:sys}
\end{figure}
%%%%%%%%%%%%%%%%%%%%%%%%%%%%%%%

These remarkable findings motivate us to reexamine the physics of triplet SCs in systems with negligible spin-orbit coupling. We focus on a particular state which is natural for TBG and related graphene-based moir\'{e} materials: a fully spin-polarized triplet superconductor, in which only electrons of one spin component (along a spontaneously chosen quantization axis) are paired. This phase is analogous to the $\mathrm{A}_1$ and $\beta$ phases discussed in the context of superfluid \Helium~\cite{vollhardt2013superfluid,Bruder1986Symmetry,volovik2003universe}. The order parameter of such a superconductor is specified by the spin direction of the condensate and its phase; topologically, the order parameter space is equivalent to the space of three-dimensional rotation matrices, $SO(3)$~\cite{Mermin1979}; see Fig.~\ref{fig:sys}\triad. 
This property
crucially determines the topology of the order parameter space, and hence
the ability of the superconductor to carry stable supercurrents. 

In a ring geometry, the superconducting phase of an ordinary (e.g., singlet) SC can wind any integer number of times around the hole of the ring. 
Mathematically, this is expressed by the homotopy group $\pi_1[U(1)]= \mathbb{Z}$. Changing the winding number requires creating a topological defect at which the magnitude of the order parameter is suppressed to zero, such as a phase slip or a vortex~\cite{halperin1979resistive,Ambegaokar1980Dynamics}. Since such defects are energetically suppressed, a configuration with a non-zero winding number (corresponding to a supercurrent) is metastable and may persist over a very long time.

In contrast, the homotopy group the order parameter of a fully spin-polarized triplet SC on a ring is $\pi_1[SO(3)]= \mathbb{Z}_2$. Physically, this means that in such a SC, two vortices can always annihilate each other, regardless of their vorticity. A configuration in which the SC phase winds twice, either around a point or in a ring geometry, can be ``untwisted'' back to a uniform configuration continuously, without diminishing the magnitude of the order parameter at any location. The untwisting process involves a temporary change in the system's magnetization; thus, the timescale for this process to occur depends on the coupling of the SC condensate to either an intrinsic or extrinsic bath with which it can exchange spin angular momentum.

The purpose of this work is to explore the unusual transport properties of fully spin-polarized triplet SCs that result from the topology of their order parameter space. We find that in such SCs, a supercurrent-carrying state is much more fragile than in an ordinary SC. The critical current density, $J_c$, that the system can sustain in a metastable state depends on the applied Zeeman field, $B$.
For $B=0$, we find that $J_c$
scales inversely with the system size along the current direction, and depends on both the spin and phase stiffnesses. A small voltage applied across the SC results in a time-dependent current with a direct-current (DC) component of magnitude close to this critical current, and an alternating-current (AC) component with a fundamental frequency $\tilde{\omega}_J = eV/\hbar$, i.e., \emph{half} of the usual Josephson frequency across a SC weak link; see Fig.~\ref{fig:sys}\joseph. This doubled periodicity is directly related to the $\mathbb{Z}_2$ topological structure of the order parameter space. For sufficiently large applied Zeeman field, the critical current scales as $J_c\propto\sqrt{B}$ independently of system size.

We demonstrate these phenomena within a time-dependent Ginzburg-Landau model, where we assume that the system is coupled to a bath with which it can exchange both energy and spin angular momentum. For currents larger than the critical current density mentioned above, $J_c$, the model provides a prediction for the current-voltage relation: $V \propto (J-J_c)^2$, up to logarithmic corrections, where $J$ is the DC component of the current. When an external Zeeman field is applied (e.g., an in-plane magnetic field in a two-dimensional system), it pins the direction of the spin magnetization, and the properties of the system rapidly cross over to those of an ordinary SC. We hope that these predictions will provide guidance to experiments in novel exotic SCs, such as TDBG, where they can be used to confirm or invalidate the existence of a fully spin-polarized triplet SC.

This paper is organized as follows. In Sec.~\ref{sec:phys} we review the properties of a fully-spin polarized SC, its distinction from other triplet SC phases, and the topology of its order parameter space. We then describe the physical picture and summarize the main results of the paper. Sec.~\ref{sec:model} describes the Ginzburg-Landau (GL) model. In Sec.~\ref{sec:landscape} we study the energy landscape of the model, either with or without an applied Zeeman field, deriving the maximum supercurrent the system can carry in a metastable state. Sec.~\ref{sec:tdgl} introduces a time-dependent extension of the GL model. This model allows us to study dynamic phenomena, such as the properties of the system in the presence of a finite applied voltage. The results are discussed in Sec.~\ref{sec:discuss}. The Appendices contain technical details of the solution of the time-dependent Ginzburg-Landau (TDGL) equations.

\section{Physical picture \& main results}\label{sec:phys}

In order to gain physical intuition of spin-polarized triplet superconductivity, we begin with a short review of the relevant order parameters. 
Those who are already familiar with this material may opt to skip to the results which are presented in Sec.~\ref{sec:results}.

\subsection{Review of triplet superconductivity and application to multi-valley systems}\label{sec:rev}

\subsubsection{Order parameter}

The order parameter of a triplet SC
can be represented in terms of the so-called d-vector, which is a complex vector defined as
$\dd_{\mathbf{k}} = \tavg{\psi^\dagger_{\mathbf{k}} i \sigma_2\vsigma\psi^\dagger_{-\mathbf{k}}}$, 
where $\psi^\dagger_{\mathbf{k}}=(\psi^\dagger_{\mathbf{k}\uparrow},\psi^\dagger_{\mathbf{k},\downarrow})$, and the Pauli matrices $\vsigma=(\sigma_1,\sigma_2,\sigma_3)$ act on the spin degrees of freedom $(\uparrow,\downarrow)$.
The Pauli principle forces $\dd_{\mathbf{k}} = -\dd_{-\mathbf{k}}$.\footnote{In a rotationally symmetric system, this implies that Cooper pairs carry an odd orbital angular momentum.}

The existence of multiple valleys in systems such as TBG and TDBG enables the possibility of a valley-singlet state which does not require the order parameter to be momentum dependent within each valley. (Exotic spin-singlet order parameters have been proposed in monolayer graphene~\cite{Uchoa2007Superconducting}; here, we focus on spin-triplet pairing.)   
Considering a system at a finite density away from charge neutrality with a finite Fermi surface, we project the pairing potential to Bloch states near the Fermi surfaces of the two valleys.

In this context, we denote
creation operators of electrons at the $\mathrm{K}_+$ and $\mathrm{K}_-$ valleys 
by the two-component spinors
$\psi^\dagger_{+}=(\psi^\dagger_{+,\uparrow},\psi^\dagger_{+,\downarrow})$ and $\psi^\dagger_{-}=(\psi^\dagger_{-,\uparrow},\psi^\dagger_{-,\downarrow})$, respectively. 
An inter-valley, valley-singlet superconducting order parameter corresponds to~\footnote{Note that taking $\vec{k}$ to $-\vec{k}$ exchanges $\mathrm{K}_+$ with $\mathrm{K}_-$.}
\begin{equation}\label{eq:ddef}
\dd=\tavg{\psi^\dagger_+ i \sigma_2\vsigma\psi^\dagger_-},
\end{equation}
where $\dd$ is independent of momentum within each valley. For simplicity, we will focus on this type of state henceforth.

The d-vector 
may be conveniently described using two real vectors $\dd^{(1,2)}$,
\begin{equation}
\dd=\dd^{(1)}+i\dd^{(2)},
\end{equation}
We define the pairing polarization $\vm$ and superfluid density $\rho$ as
\begin{align}\label{eq:mrho}
&\vm =\frac{1}{2i}{\dd}^{\ast}\times{\dd}=\dd^{(1)}\times\dd^{(2)}, &\rho={\dd}^{\ast}\cdot{\dd},
\end{align}
see Fig.~\ref{fig:sys}\triad. 
At temperatures far below the mean-field superconducting phase transition, we may treat the problem within the London limit where the magnitude of the order parameter is essentially fixed,
\begin{equation}
\dd^\ast\cdot\dd=2d_0^2.
\end{equation}

Note that the pairing polarization $\vec{m}$ arises from the anomalous expectation value $\vec{d}$ in Eq.~(\ref{eq:ddef}).
At equilibrium, the pairing polarization is
coupled to the physical magnetization (spin polarization) density, and proportional to it. This relation also holds for slowly varying magnetization textures. Thus, below we will discuss the behavior of the pairing polarization interchangeably with that of the magnetization density.

%%%%%%%%%%%%%%%%%%%%%%%%%%%%%%%
\begin{figure*}[t]
    \centering
    \includegraphics[width=1\linewidth]{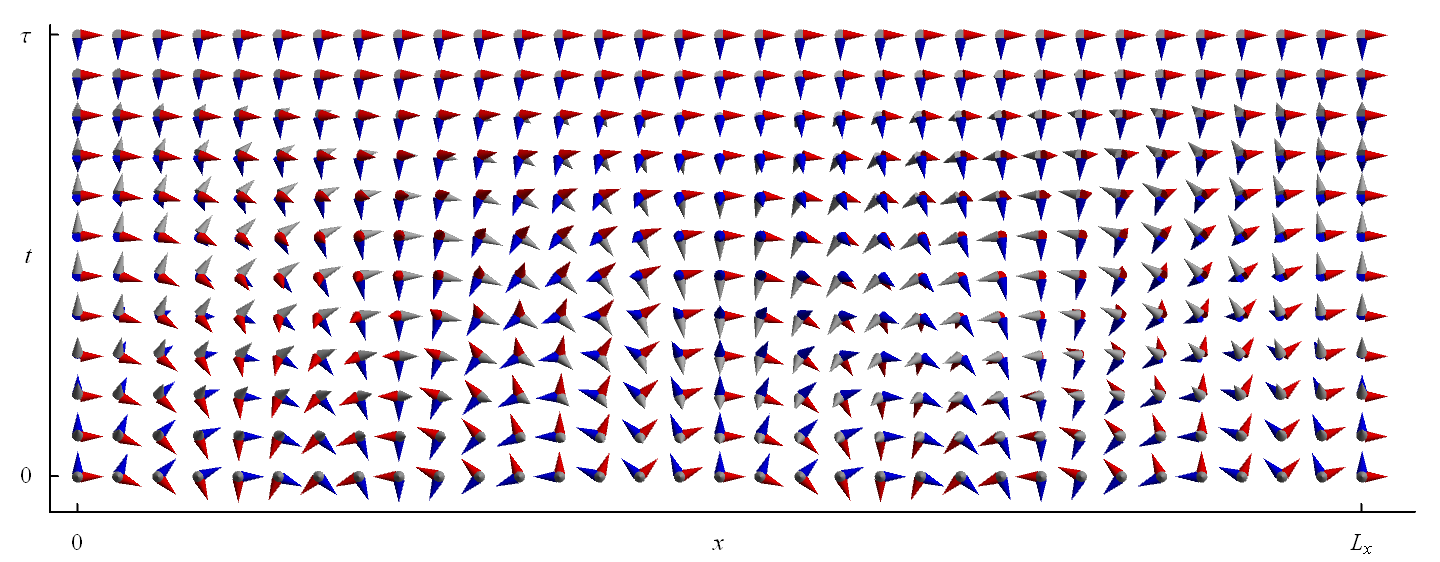}
    \caption{Continuous trajectory for supercurrent relaxation. We present the trajectory of Eq.~(\ref{eq:traj}) relaxing a supercurrent of $J=2J_0$ at $t=0$ to a supercurrent of $J=0$ at $t=\tau$. The order parameter components, $\dd^{(1)}$ and $\dd^{(2)}$, are depicted by the red and blue arrows, the gray arrows depict the pairing polarization, $\vm$; see Fig.~\ref{fig:sys}\triad. }
    \label{fig:traj}
\end{figure*}
%%%%%%%%%%%%%%%%%%%%%%%%%%%%%%%

\subsubsection{Symmetry classification of valley-singlet states}\label{sec:sym}
Triplet superconducting phases have been studied in great detail in the context of superfluid \Helium; see Refs.~\onlinecite{vollhardt2013superfluid,volovik2003universe} for a review.
These works naturally focused on p-wave phases. However, as discussed above, in multi-valley materials, there is a possibility of a valley-singlet phase. Nevertheless, a similar symmetry classification of the possible phases can be carried out.

Neglecting spin-orbit coupling, the free energy of the system is invariant under transformations in the $SO(3)_{S}$ group of global spin rotations and in the group of $U(1)_{\phi}$ gauge transformations.
We thus distinguish three distinct triplet phases, according to their patterns of broken symmetries:

\begin{itemize}

\item[(i)] {\it Non-spin-polarized, unitary “spin nematic” phase.}-- In this phase, the pairing polarization vanishes, $\vm=0$, corresponding to
$\dd^{(1)}\,||\,\dd^{(2)}$. The symmetry is reduced to the group of $SO(2)_{S}$ spin rotations around a preferred direction and the $[\mathbb{Z}_2]_{S+\phi}$ operation of a simultaneous $\pi$ spin rotation around an axis perpendicular to $\dd_{1,2}$ and a $\pi$ gauge transformation. The order parameter degeneracy space is given by
\begin{equation}
\frac{SO(3)_{S}\times U(1)_{\phi}}{SO(2)_{S}\times[\mathbb{Z}_2]_{S+\phi}}=[S^2/\mathbb{Z}_2]_{S}\times U(1)_{\phi}.
\end{equation}
This phase has the same pattern of spin and gauge symmetry breaking as the $\mathrm{A}$ and $\mathrm{B}_2$ phases of \Helium~\cite{vollhardt2013superfluid,Bruder1986Symmetry}. We note, however, that the phases of \Helium differ by the breaking of orbital symmetry which is absent for the valley-singlet phase described here.

\item[(ii)] {\it Fully spin-polarized non-unitary triplet.}-- In this phase $\dd\cdot\dd=0$, corresponding to $\dd^{(1)}\perp\dd^{(2)}$ and $|\dd^{(1)}|=|\dd^{(2)}|$. 
The condition $\dd \cdot \dd = 0$ can be interpreted as the vanishing of a charge $4e$ scalar order parameter $\Delta_{4e}=\dd\cdot\dd$.
The symmetry is reduced to that of $U(1)_{S+\phi}$ rotations around the pairing polarization direction, $\vm$, matched with gauge transformations by the same phase.
The order parameter degeneracy space is given by
\begin{equation}
\frac{SO(3)_{S}\times U(1)_{\phi}}{U(1)_{S+\phi}}=SO(3)_{S+\phi}.
\end{equation}
This situation is similar to the $\mathrm{A}_1$ and $\beta$ phases of \Helium~\cite{vollhardt2013superfluid,Bruder1986Symmetry}, up to orbital symmetries (see above).

\item[(iii)] {\it Partially spin-polarized non-unitary triplet.}-- In this phase neither $\vm$ nor $\Delta_{4e}$ vanish. The symmetry is completely broken and the degeneracy space is given by
\begin{equation}
SO(3)_{S}\times U(1)_{\phi}.
\end{equation}
This situation is similar to the $\mathrm{B}$ and $\mathrm{A}_2$ phases of \Helium~\cite{vollhardt2013superfluid,Bruder1986Symmetry}, up to orbital symmetries (see above).

\end{itemize}

Recent theoretical works~\cite{lee2019theory,scheurer2019pairing} have pointed out that the order parameter of TDBG might be in a fully spin-polarized state, at least in a part of the phase diagram. Motivated by this possibility, together with a fundamental interest in the novel properties that we uncover below, in the remainder of this work we will focus on the properties of fully spin-polarized triplet SCs [phase (ii) above].

\subsection{Supercurrent decay mechanisms}\label{sec:results}

The robustness of persistent supercurrents in ordinary SCs has a topological origin: in a superconducting ring, the winding number of the phase of the order parameter around the ring is an integer that cannot be changed by a small, local perturbation.
For fully spin-polarized triplet SCs, the $SO(3)$ topology of the order parameter configuration space~\footnote{The existence of $\mathbb{Z}_2$ defects has been studied in the context of superfluid \Helium; see Ref.~\onlinecite{vollhardt2013superfluid}.},
\begin{equation}\label{eq:topo}
\pi_1[SO(3)]=\mathbb{Z}_2,
\end{equation}
has important consequences for the stability of the supercurrent. Specifically, consider an initial current-carrying configuration with a constant pairing polarization, and a certain number $n$ of windings of the superconducting phase across the system. A continuous relaxation event, such as the one illustrated in Fig.~\ref{fig:traj}, can reduce the phase winding to $n-\delta$, for some even integer, $\delta$. The unwinding event may involve a finite free energy barrier,
which depends on the initial number of windings, $n$, on the ratio between the phase and pairing polarization stiffnesses, and on the applied Zeeman field. The barrier originates from the fact that the unwinding process involves twisting the pairing polarization in space and changing its average direction. 

Throughout this paper we assume the presence of sufficiently large fluctuations such that the system explores its configuration space and that any energetically favorable configuration (without an intermediate energy barrier) is quickly explored; we discuss relaxing this assumption at Sec.~\ref{sec:mag}.
We analyze the energy landscape of the continuous unwinding process in Sec.~\ref{sec:landscape}. The result is that the energy barrier vanishes above a certain current density, which we define as the critical current $J_c$. The critical current has two contributions,
\begin{equation}\label{eq:main}
J_{c}\sim\begin{cases}
e\sqrt{k_B T_{\mathrm{BKT}}\mu B} & \mu B \gg {k_B T_{\mathrm{BKT}}}/{L_x^2}, \\
{e k_B T_{\mathrm{BKT}}}/{L_x} & \mu B \ll {k_B T_{\mathrm{BKT}}}/{L_x^2}.
\end{cases}
\end{equation}
where $\mu$ is the magnetic moment density in the material, $L_x$ is the length of the sample along the current direction, $B$ is the applied Zeeman field, and $T_{\mathrm{BKT}}$ is the Berezinskii-Kosterlitz-Thouless (BKT) transition temperature.
Intuitively, in for a small Zeeman field, we find only a mesoscopic contribution that scales inversely with system size, corresponding to a finite number of windings. 
On the contrary, a large Zeeman field acts to lock the pairing polarization, increasing the critical current. 
For a large enough Zeeman field, $J_c$ exceeds the intrinsic (microscopic) critical current density; continuous unwinding events then become irrelevant, and the system behaves as an ordinary SC. 

A direct physical manifestation of the order parameter topology can be observed when a voltage is applied across a fully spin-polarized SC [Fig.~\ref{fig:sys}\sys]. The voltage causes the superconducting phase at the right lead to wind relative to the phase at the left lead, at a constant rate $\omega_J = 2 eV/\hbar$. Once the phase difference across the system has increased by $4\pi$, there is no topological obstruction to continuously ``unwind'' the phase twist, relaxing the supercurrent. After the unwinding event, the phase difference (and hence the supercurrent) starts growing linearly in time again.
Assuming that the dissipation rate is large compared to the Josephson frequency (a condition to be discussed further in Sec.~\ref{sec:vortex}), we find that the current undergoes periodic oscillations with frequency $\tilde{\omega}_J = 2eV/\hbar$, i.e., \emph{half} of the ordinary AC Josephson frequency [see Fig.~\ref{fig:sys}\joseph]. In fact, we find that at large magnetic fields it is energetically favorable to unwind more than two windings at once thus further \emph{fractionalizing} the ordinary AC Josephson frequency, i.e., $\tilde{\omega}_J = 2eV/\delta\hbar$ with $\delta\ge2$ an even integer. This is one of our main results. We substantiate it by solving a time-dependent Ginzburg-Landau model in Sec.~\ref{sec:tdgl}.

\section{Ginzburg-Landau model}\label{sec:model}

\subsection{Complex vector description}

The starting point of our analysis of a spin-polarized triplet SC is most generic gauge-invariant form of the  of the free energy, expanded in long wavelengths up to second order in derivatives, as a function of the complex d-vector order parameter:
\begin{align}\label{eq:Fd}
F[\dd] =\int d^{2}r\bigg\{ & \frac{\kappa_d}{2d_{0}^{2}}|(\nabla-iq\A)\dd|^{2}+\frac{\kappa_m}{2d_{0}^{4}}\left(\nabla(\tfrac{1}{2i}{\dd}^{\ast}\times{\dd})\right)^{2}\nonumber\\
& -\frac{\mu}{d_0^2}\B\cdot(\tfrac{1}{2i}{\dd}^{\ast}\times{\dd})+U(\dd)\bigg\} .
\end{align}
Here, $\kappa_d$ is a 
generalized phase stiffness, $\kappa_m$ is the {excess} spin stiffness, $\B$ is the applied Zeeman field, $\mu$ is the magnetic moment density, and the potential $U(\dd)$ can be any gauge-invariant function\footnote{Note that the term proportional to $\kappa_d$ itself gives an energy cost for twists of the pairing polarization; therefore $\kappa_m$ captures the {\it excess} spin stiffness beyond that included in the generalized phase stiffness term.}. In order for the free energy to be bounded from below,
the generalized phase stiffness, $\kappa_d$, and the excess spin stiffness, $\kappa_m$, must satisfy $\kappa_d\ge 0$ and $\kappa_m\ge -\kappa_d$.
We have opted for the Weyl-gauge, i.e., zero electric scalar potential, $\phi=0$, such that the system is coupled to the magnetic vector potential $\A$; for generality, we keep the charge, $q$, generic (in electronic SCs $q=-2e$).  

The Poisson bracket between the components of $\dd$ would be useful is the study of the dynamics (Sec.~\ref{sec:tdgl}),
\begin{equation}
\{d_i(\rr)^\ast,d_j(\rr')\} =2i\delta_{ij}\delta(\rr-\rr').
\end{equation}
These give rise to the following relations: 
\begin{align}
\{m_{i}(\rr),m_{j}(\rr')\} & =\varepsilon_{ijk}m_{k}(\rr)\delta(\rr-\rr'),\nonumber\\
\{d_{i}(\rr),m_{j}(\rr')\} & =\varepsilon_{ijk}d_{k}(\rr)\delta(\rr-\rr'),\nonumber\\
\{\dd(\rr),\rho(\rr')\} & =-2i \dd(\rr)\delta(\rr-\rr'),
\end{align}
consistent with $\mathbf{m}$ and $\rho$ being the generators of spin rotations and gauge transformations, respectively.

The supercurrent density is given by
\begin{equation}\label{eq:Jd}
\J=-\frac{\delta F}{\delta\A}=q\kappa_d\frac{{\dd}^{\ast}\cdot\nabla\dd-\dd\cdot\nabla{\dd}^{\ast}}{2id_0^2}-q^{2}\kappa_d \frac{\dd^{\ast}\cdot\dd}{d_0^2}\A.
\end{equation}

\subsection{Unitary matrix description}\label{sec:uni}

A very useful description of the $SO(3)$ topology for the purpose of explicit computations is through its $SU(2)$ double cover. This description is applicable  only for the fully spin-polarized phase (see Sec.~\ref{sec:sym}), where it captures the three-dimensional nature of the degeneracy space. For every matrix $u\in SU(2)$ we can associate a complex d-vector
\begin{equation}
{\dd}=d_{0}\frac{1}{2}\Tr{u(\sigma_{1}+i\sigma_{2})u^{\dagger}\bm{\sigma}}.\label{eq:u}
\end{equation}
The double cover is evident by the identical order parameters corresponding to both $u$ and $-u$. The homotopy expressed in  Eq.~(\ref{eq:topo}) follows directly from this property (see discussion in Sec.~\ref{sec:landscape}). 
This description holds in the London limit where $|\dd|^2=2d_0^2$. 
Gauge transformations are implemented as
\begin{equation}
\A\mapsto\A+\nabla\Lambda, \qquad \dd\mapsto e^{iq\Lambda}\dd, \qquad u\mapsto ue^{i\frac{q}{2}\Lambda\sigma_3},
\end{equation}
where $\Lambda(\vec{r})$ is a scalar function.
On the other hand, a spin rotation
of angle $\theta$ around an axis $\hat{\mathbf{n}}$ acts by
\begin{equation}
u\mapsto e^{i\frac{\theta}{2}\hat{\mathbf{n}}\cdot\bm{\sigma}}u.
\end{equation}
The most general form of the free energy invariant under gauge and spin rotation transformations [Eq.~\eqref{eq:Fd}] is given by
\begin{align}\label{eq:Fu}
F[u] =&\int d^{2}r\bigg\{ \frac{\kappa_m-\kappa_d}{4}\Tr{\nabla(u\sigma_3u^\dagger)\cdot\nabla(u\sigma_3u^\dagger)}\nonumber\\
&+2\kappa_d\Tr{\left(\nabla u^\dagger+i\tfrac{q}{2}\A\sigma_3u^\dagger\right)\left(\nabla u-i\tfrac{q}{2}\A u\sigma_3\right)}\nonumber\\
&-\frac{\mu}{2}\B\cdot\Tr{u\sigma_3u^\dagger\vsigma}\bigg\}.
\end{align}
This unitary matrix description automatically enforces both $\dd^\ast\cdot\dd=2d_0^2$ and $|\dd\cdot\dd|^2=0$, hence, we have dropped the constant scalar potential, $U(\dd)=const.$.
Within this description, the supercurrent density [Eq.~(\ref{eq:Jd})] and the pairing polarization [Eq.~\eqref{eq:mrho}] are given by
\begin{align}\label{eq:Ju}
\J &= -\frac{\delta F}{\delta\A}=
-iq\kappa_d\Tr{u^\dagger\nabla u\sigma_3- u\sigma_3\nabla u^\dagger }-2q^2\kappa_d\A, \nonumber\\
\vm &= d_0^2\frac{1}{2}\Tr{u\sigma_3u^\dagger\vsigma}.
\end{align}

\section{Energy landscape}\label{sec:landscape}

We now consider the energetics of the continuous unwinding process for supercurrent relaxation. In general, a supercurrent carrying configuration would be energetically less favorable than a configuration with a smaller supercurrent. This less favorable configuration may be either a metastable or an unstable state. As long as the higher-energy configuration is metastable, either thermally activated or quantum tunneling events would act to relax the current; however, such processes are expected to be exponentially suppressed. Therefore, as the supercurrent gradually increases (e.g., in response to an applied DC voltage; see Fig.~\ref{fig:sys}), the system should stabilize on a value of the current close to its lowest unstable state.

\subsection{Linear stability}

We seek the lowest possible supercurrent for which a supercurrent carrying state ceases to be metastable (i.e., where an instability develops).

We consider an applied Zeeman field in the plane of the SC, hence, without loss of generality, we set $\B=B\hat{\mathbf{x}}$. In the presence of this Zeeman field, the system always has as a fixed point, $\delta F/\delta \dd=0$,
where the d-vector exhibits a uniform configuration described by
\begin{equation}\label{eq:1i0}
\dd_0=(0,1,i)d_0.
\end{equation}
This configuration may be set to carry arbitrarily high currents by applying a vector potential $\A = -\frac{1}{2q^2\kappa_d}\J$, with a current $\J=J\hat{\vec{x}}$; see Eq.~\eqref{eq:Jd}.
In search of an instability, we study small deviations, $\dd=\dd_0+\delta\dd$, from the
configuration $\dd_0$ in Eq.~(\ref{eq:1i0}), with
\begin{align}
\delta\dd^{(1)}/d_0&=(-\eta_{2},\ -\tfrac{1}{2}(\eta_{2}^{2}+\eta_{3}^{2}),\ \eta_{3}),\nonumber\\
\delta\dd^{(2)}/d_0&=(\eta_{1},\ -\eta_{3},\ -\tfrac{1}{2}(\eta_{1}^{2}+\eta_{3}^{2})),\\
\delta(\dd^{(1)}\times\dd^{(2)})/d_0^2 &=(-\tfrac{1}{2}(\eta_{1}^{2}+\eta_{2}^{2}),\ \eta_{2},\ -\eta_{1}),
\end{align}
where $\bm{\eta} = (\eta_{1},\eta_2, \eta_3)$ is real, and $|\bm{\eta}| \ll 1$. 
Here we have taken the most general deviation respecting the fully spin-polarized $SO(3)$ structure, $\dd^\ast\cdot\dd=2d_0^2$, and $\dd\cdot\dd=0$, up to second order in $\bm{\eta}$. Expanding the free energy 
to second order in the deviation $\bm{\eta}$ yields
\begin{align}
\delta F[\bm{\eta}]&=\int d^{2}r\bigg\{\frac{\kappa_m+\kappa_d}{2}\left[(\nabla\eta_{1})^{2}+(\nabla\eta_{2})^{2}\right]+\kappa_d(\nabla\eta_{3})^{2} \nonumber\\
&-\frac{1}{2q}\J\cdot\left(\eta_{1}\nabla\eta_{2}-\eta_{2}\nabla\eta_{1}\right)+\frac{\mu B}{2}(\eta_{1}^{2}+\eta_{2}^{2})\bigg\}.
\end{align}
We seek the least stable direction, defined with respect to the curvature of $\delta F$, and find that it takes the form
\begin{align}
\bm{\eta}&=(\eta\cos kx,\ \eta\sin kx,\ 0).
\end{align}
For such a deviation, the corresponding change of the free energy is given by
\begin{align}\label{eq:quadeta}
\delta F[{\eta}]&=\int d^2 r \tfrac{\eta^{2}}{2}\left[(\kappa_{m}+\kappa_{d})k^{2}-\tfrac{1}{q}Jk+\mu B\right].
\end{align}
The wavenumber, $k$, for which the free energy is minimal, is thus $k=\frac{J}{2q(\kappa_m+\kappa_d)}$. However, mesoscopic finite size effects must be taken into consideration. Specifically, a system on a cylinder of circumference $L_x$ cannot support fluctuations smaller than the fundamental wavelength, i.e.,
\begin{equation}
k\simeq\max\left\{\frac{2\pi}{L_x},\frac{J}{2q(\kappa_m+\kappa_d)}\right\},
\end{equation}
where strictly speaking, the wavenumber, $k$, must take the closest positive integer multiple of $\frac{2\pi}{L_x}$. 
We thus find the critical current for the onset of instability, $J_c$,
as the value of $J$ for which the curvature of the quadratic dependence of $\delta F$ on $\eta$ [Eq.~\eqref{eq:quadeta}] becomes negative:
\begin{equation}\label{eq:Jc1}
J_{c}\simeq\begin{cases}
2q\sqrt{(\kappa_m+\kappa_d)\mu B} & \mu B \ge \frac{4\pi^2(\kappa_m+\kappa_d)}{L_x^2}, \\
\frac{2\pi q(\kappa_m+\kappa_d)}{L_x}+\frac{qL_x \mu B}{2\pi} & \mu B < \frac{4\pi^2(\kappa_m+\kappa_d)}{L_x^2}.
\end{cases}
\end{equation}
For $J\ge J_c$, the uniform pairing-polarization fixed-point configuration in Eq.~\eqref{eq:1i0} becomes a saddle point, and a spontaneous polarization texture should form.
Nevertheless, this does not guarantee the existence of a trajectory in configuration space which connects the system to a new stable fixed-point without traversing an energy barrier. In the next section, we thus estimate the possible energy barriers and seek out these trajectories.

\subsection{Unwinding trajectories and energy barriers}

We move on to estimate the free energy barrier for relaxing two windings of the superconducting phase continuously, as allowed by the topology of the order parameter [Eq.~(\ref{eq:topo})]. 

This is easiest to describe using the unitary matrix presentation. Moreover, as the only applied electromagnetic field is the in-plane Zeeman field, we opt to pick the gauge of $\A=0$ at all positions within the plane of the SC. In this gauge and presentation, a uniform persistent supercurrent carrying state is given by
\begin{equation}\label{eq:uJ}
u(\rr)=u_{0}e^{i\sigma_{3}\frac{Jx}{4q\kappa_d}},
\end{equation}
where $u_0$ is any $SU(2)$ matrix. We consider a system with cylindrical geometry of height $L_y$ and circumference $L_x$. The order parameter, $\dd$, thus has periodic boundary conditions along $x$, which force either periodic or antiperiodic boundary conditions for $u$. These correspond to the two $\mathbb{Z}_2$ classes of maps from the cylinder to the order parameter space; see Eq.~\eqref{eq:topo} and discussion in Sec.~\ref{sec:uni}. 
As a consequence, the supercurrent is quantized to $J=J_{0}n$, where $n\in\mathbb{Z}$ and
\begin{equation}\label{eq:J0}
J_{0}=\frac{4\pi q\kappa_d}{L_x}
\end{equation}
is the fundamental current; this configuration is overlaid in Fig.~\ref{fig:sys}\sys.

From the $\mathbb{Z}_2$ order parameter topology, we know that for any \emph{even} integer, $\delta$, there exist trajectories  $u(\rr,t)$ with $n$ windings at $t=0$ and $n-\delta$ windings at some later time $t=\tau$, in which the order parameter magnitude remains constant throughout the deformation.
Among these, we seek the trajectory with the minimal energy barrier.

Since $\B=B\hat{\mathbf{x}}$, 
we look for trajectories with uniform pairing polarization $\vm=d_0\hat{\mathbf{x}}$ at initial and final times $t=0$ and $t = \tau$, respectively: 
\begin{align}
u(\rr,t=0) & = u_0 e^{i\pi\sigma_{3}n\frac{x}{L_x}},\label{eq:traj2}\\
u(\rr,t=\tau) & = u_0 e^{i\pi\sigma_{3}(n-\delta)\frac{x}{L_x}},\label{eq:traj3}
\end{align}
with $u_0=e^{-i\frac{\pi}{4}\sigma_2}$.
We study the following variational  family of trajectories $u(\rr,s_1,s_2)$, which connect these initial and final states, and parametrically depend on time via 
$s_{1,2}(t=0)=0$ and $s_{1,2}(t=\tau)=1$: 
\begin{equation}
u(\rr,s_1,s_2)  = u_0 e^{-i\frac{\pi}{2}\sigma_{1}s_2}e^{i\pi\sigma_{3}\tfrac{1}{2}\delta\frac{x}{L_x}}e^{i\frac{\pi}{2}\sigma_{1}s_1}e^{i\pi\sigma_{3}(n-\tfrac{1}{2}\delta)\frac{x}{L_x}}.\label{eq:traj} 
\end{equation}
Comparing Eqs.~(\ref{eq:traj2})-(\ref{eq:traj}) 
with Eq.~(\ref{eq:uJ}), 
it is evident that this trajectory connects a supercurrent-carrying state with $n$ windings to a state with $n-\delta$ windings. For example, the trajectory $s_1=t/\tau$, $s_2=0$ is shown in Fig.~\ref{fig:traj}.

By inserting the ansatz for $u(\rr,t)$ in Eq.~(\ref{eq:traj}) into Eq.~\eqref{eq:Ju}, one finds that the variation of $s_1$ (for $s_2$ held fixed at 0) acts to dissipate the current but flips the pairing polarization relative to its initial orientation parallel to the external field.
The variation of $s_2$ acts to re-align the pairing polarization. 
An upper bound on the free energy barrier can be obtained by considering an example trajectory where $s_1$ rises from 0 to 1 
strictly before $s_2$ does so. Such a trajectory has the appealing property that the free energy density 
is spatially homogeneous, and is given by
\begin{align}
& 
\tfrac{1}{L_xL_y}F[u(\rr,s_1, 0)]=-\mu B \cos(\pi s_1)+\tfrac{4\pi^{2}}{L_x^2}\Big[\kappa_d n^{2} \nonumber\\
&-\left(\kappa_d(2n\delta-\delta^2)-(\kappa_m-\kappa_d)\tfrac{1}{2}\delta^2\cos^{2}(\tfrac{\pi s_1}{2})\right)\sin^{2}(\tfrac{\pi s_1}{2})\Big],\nonumber \\
& 
\tfrac{1}{L_xL_y}F[u(\rr,1, s_2)]=\mu B \cos(\pi s_2)+\tfrac{4\pi^{2}}{L_x^2}\kappa_d (n-\delta)^{2}.
\end{align}
This enables us to obtain the maximal energy barrier along this trajectory,
\begin{equation}\label{eq:barrier}
\Delta F[u]\defeq\underset{0\le t_1\le t_2\le\tau}{\max}\left\{F[u(\rr,t_2)]-F[u(\rr,t_1)]\right\},
\end{equation}
which provides an upper bound on the energy barrier amongst all possible trajectories, $\Delta F\defeq\min_u\{\Delta F[u]\}$.

Beyond a certain critical current, there is no energy barrier for supercurrent relaxation. 
Through the dependence of the current on the number of windings, $J=J_0n$, each variational trajectory yields an upper bound $J_c\le J_c^\mathrm{var}(\delta)$ at which the free energy barrier for the continuous unwinding process vanishes. By optimizing over the number of unwindings $\delta$, we find the best variational estimate, $J_c^\mathrm{var}\defeq\min_\delta\{J_c^\mathrm{var}(\delta)\}$.

Particularly, for $\kappa_m>\kappa_d$, our variational trajectories turn out to provide a strict estimate:
\begin{align}\label{eq:Jc2var}
J_c^\mathrm{var}(\delta)&=\tfrac{\pi q(\kappa_m+\kappa_d)\delta}{L_x}+\tfrac{qL_x \mu B}{\pi\delta},\\
J_c^\mathrm{var}&\defeq\min\nolimits_\delta\{J_c^\mathrm{var}(\delta)\}=J_c.
\end{align}
Remarkably, $J_c^\mathrm{var}$ \emph{equals} $J_c$ found by our linear stability analysis [Eq.~\eqref{eq:Jc1}].
This also provides us with an expression for the optimal number of simultaneous unwindings,
\begin{equation}\label{eq:period}
\delta \simeq \frac{L_x}{\pi}\sqrt{\frac{\mu B}{\kappa_m+\kappa_d}},
\end{equation}
where strictly speaking, $\delta$, must be the closest positive even integer.
As discussed in Sec.~\ref{sec:results}, the number of simultaneous unwindings, $\delta$, determines the periodicity of the AC Josephson effect in our system, $\tilde{\omega}_J = qV/\delta\hbar$.

At currents close to the critical current, $J\sim J_c$, the free energy barrier in Eq.~(\ref{eq:barrier}) satisfies:
\begin{equation}\label{eq:barrierJc}
\Delta F\le
\begin{cases}
0 & J\ge J_c,\\
\frac{L_x L_y}{2q^2(\kappa_m-\kappa_d)}(J-J_c)^2 & J<J_c.
\end{cases}
\end{equation}
For $J<J_c$, current relaxation by continuous unwinding processes is possible, but requires
traversing a barrier. 
In this case, relaxation is expected to be much slower than relaxation for $J\ge J_c$, especially at low temperatures. 

Note, that for $\kappa_m\le\kappa_d$, we must consider trajectories where $s_{1,2}$ in Eq.~\eqref{eq:traj} are varied simultaneously and the variational bounds we obtain are less strict, i.e., $J_c\lneq J_c^{\mathrm{var}}$. Nevertheless, these bounds portray the same functional dependence on $B$ and $L_x$ as Eq.~\eqref{eq:Jc1} and are thus omitted for simplicity's sake.
Particularly, for $B=0$ we still find that $J_c^\mathrm{var}$ \emph{equals} $J_c$.

Moreover, as we show in Sec.~\ref{sec:tdgl}, using dynamical simulations, Eq.~(\ref{eq:traj}) provides a good approximation for the continuous unwinding trajectories that naturally emerge from the dynamics of the system, 
and the critical current is indeed well captured by Eq.~(\ref{eq:Jc1}).

\subsection{Competing mechanisms}

Throughout this analysis, we have assumed that the order parameter is uniform in the $y$ direction. 
There also exist trajectories where the unwinding process occurs in a ``domain wall'' whose width in the $y$ direction is smaller than $L_y$. The domain wall then propagates along $y$ and unwinds the phase twist in the entire system. 
However, the optimal width of the domain wall is proportional to $L_x$, and thus its formation is only favorable when $L_y$ is sufficiently larger than $L_x$. 

Another competing mechanism for current relaxation is by vortex motion perpendicular to the current direction. As discussed in Sec.~\ref{sec:discuss}, we find the continuous unwinding mechanism is dominant at sufficiently low currents and low temperatures. 

\section{Dynamic transport}\label{sec:tdgl}
So far, we have explored current relaxation from an energetic point of view. 
In this section, we support the predictions from the analysis of Sec.~\ref{sec:landscape} using an unbiased model for the order parameter dynamics that enables the system to explore its configuration space. 

\subsection{Time dependent model}
To capture the dynamics of the current relaxation process, we use a Time-Dependent Ginzburg-Landau (TDGL) formulation supplemented by a stochastic noise term~\cite{Hohenberg1977,chaikin1995principles,risken1996fokker}. Crucially for the continuous unwinding process, we must include a mechanism for the system to exchange magnetization with its environment (relaxing the constraint of angular momentum conservation).  
To this end, we assume that the system is coupled to an external spin bath; for further discussion see Sec.~\ref{sec:mag}.

Within the stochastic TDGL approach, the order parameter evolves according to a Langevin-type equation of motion of the form
\begin{equation}\label{eq:TDGLindex}
\frac{\partial d_{i}^{(\alpha)}}{\partial t}=\varepsilon^{\alpha\beta}\frac{\delta F}{\delta d_{i}^{(\beta)}} -\Gamma g_{ik}^{\alpha}g_{jk}^{\beta}\frac{\delta F}{\delta d_{j}^{(\beta)}}+g_{ij}^{\alpha}\zeta_{j}.
\end{equation}
Here, we have utilized the Einstein summation convention, with $\alpha,\beta\in\{1,2\}$ and $i,j,k\in\{1,2,3\}$. 
The three terms on the right-hand side of Eq.~(\ref{eq:TDGLindex}) are the kinetic (dissipationless) term, the dissipative term, and the source (noise) term, respectively.
We take $\zeta_j(\vec{r},t)$ to be a real Gaussian-distributed stochastic field with zero mean and correlation function
\begin{equation}
\avg{\zeta_i(\rr,t)\zeta_j(\rr',t')} =2k_{B}T\Gamma\delta_{ij}\delta(\rr-\rr')\delta(t-t').
\end{equation}
We choose $g^{\alpha}_{ik}(\dd) = \varepsilon_{ijk} d^{\alpha}_j$, which corresponds to a spin bath acting as a spatiotemporally-fluctuating
Zeeman field $\bm{\zeta}$ that induces precession of the order parameter. The form of Eq.~(\ref{eq:TDGLindex}) guarantees that the fluctuation-dissipation theorem is satisfied at thermal equilibrium; see Appendix~\ref{app:FDT}. 

In terms of the complex d-vector, The Langevin equation takes the form
\begin{align}\label{eq:TDGLcross}
\frac{\partial\dd}{\partial t}= \ &\frac{1}{i}\left(\frac{\delta F}{\delta\dd^\ast}-\frac{1}{\dd^{\ast}\cdot\dd}\left({\dd}\cdot\frac{\delta F}{\delta{\dd}^{\ast}}\right){\dd}^{\ast}\right) \nonumber\\
&-\Gamma\bigg[\left(\dd\cdot\dd^{\ast}\right)\frac{\delta F}{\delta{\dd}^{\ast}}+\left({\dd}\cdot{\dd}\right)\frac{\delta F}{\delta{\dd}} \nonumber\\
&\phantom{-\Gamma\bigg[}-\left({\dd}\cdot\frac{\delta F}{\delta{\dd}^{\ast}}\right){\dd}^{\ast}-\left({\dd}\cdot\frac{\delta F}{\delta{\dd}}\right){\dd}\bigg]+{\dd}\times\bm{\zeta}.
\end{align}
As our system is in the fully spin-polarized state, the potential in the free energy [Eq.~(\ref{eq:Fd})] must include a polarizing term, $U(\dd)=\lambda\left|\dd\cdot\dd\right|^2+\ldots$, with large coupling $\lambda$. Taking the limit $\lambda\to\infty$ generates 
the correction to the kinetic term in the first line of Eq.~(\ref{eq:TDGLcross}).

Importantly, Eq.~(\ref{eq:TDGLcross}) is manifestly gauge invariant, and, as constructed, leads to the conservation laws:
\begin{align}
&\frac{\partial (\dd\cdot\dd)}{\partial t}=0,\\
&\left.\frac{\partial (\dd^\ast\cdot\dd)}{\partial t}\right|_{\dd\cdot\dd=0}=\frac{1}{i}\left(\dd^\ast\cdot\frac{\delta F}{\delta\dd^\ast}-\dd\cdot\frac{\delta F}{\delta\dd}\right)=-\nabla\cdot\J.
\end{align}
By design, the pairing polarization $\vm =\frac{1}{2i}{\dd}^{\ast}\times{\dd}$, which is proportional to the spin magnetization at equilibrium, is not conserved.

\subsection{Solutions}\label{sec:sol}

To compute transport in the system, we consider a cylindrical geometry with circumference $L_x$ and height $L_y$, where a current flows due to the presence of an applied electromotive force, equivalent to a DC voltage, $V$.
Within our stochastic model, we thus seek the mean current, $\avg{J(t)}_{\bm{\zeta}}$ (averaged over all realizations of the noise, $\bm{\zeta}$), 
that flows due to a uniform, constant electric field
\begin{equation}
\A(\rr,t)=-\E t.
\end{equation}
Here, $\vec{E} = V \hat{\vec{x}}/L_x$.
Since the TDGL equations [Eq.~(\ref{eq:TDGLcross})] are nonlinear stochastic partial differential equations, their solutions are cumbersome to write. Therefore, we leave the details of the solution to Appendices \ref{app:analytic} and \ref{app:numeric} and present here the main results.

We study the dependence on 
the model parameters, {i.e.}, the stiffnesses $\kappa_d$ and $\kappa_m$, sample dimensions $L_x$ and $L_y$, fluctuation coefficient $\Gamma$, temperature $T$, and applied voltage $V$.
We focus on certain analytically tractable limits of physical interest.
First, to avoid the domain-wall formation discussed in Sec.~\ref{sec:landscape}, we keep $L_y$ sufficiently small compared to $L_x$. Current relaxation by transverse motion of vortices is neglected, assuming that the temperature is sufficiently low (see discussion in Sec.~\ref{sec:vortex}). Under these conditions, the order parameter may be assumed to be independent of $y$. Second, for simplicity, we study the case of large fluctuation and dissipation, $\Gamma\gg d_0^{-2}$, such that the kinetic term in Eq.~(\ref{eq:TDGLcross}) is negligible compared to the dissipation and noise terms. The applied Zeeman field is moreover set to zero. Finally, in order to work in the regime where relaxation occurs through well-separated-in-time individual unwinding events, we focus on low temperatures such that $k_{B}T\ll L_y\sqrt{\kappa_d qV/\Gamma}$; see Eq.~(\ref{eq:lowtemp}) in Appendix~\ref{app:analytic}.

Below, we present the time dependent solution of the TDGL equations, Eq.~(\ref{eq:TDGLcross}), as well as the the long-time-averaged mean supercurrent,
\begin{equation}\label{eq:Javg}
\bar{J}=\lim_{\tau\to\infty}\frac{1}{\tau}\int_{0}^{\tau}dt\,\avg{J(t)}_{\bm{\zeta}}.
\end{equation}
Note that, as the system is self averaging, 
the long-time averaged current takes the value $\bar{J}$ for each generic realization, $\bar{J}=\lim_{\tau\to\infty}\frac{1}{\tau}\int_{0}^{\tau}dt\, J(t)$.

\subsubsection{Energy landscape predictions}\label{sec:pred}

Our analysis of the continuous supercurrent dissipation mechanism and estimates of the critical supercurrent in Sec.~\ref{sec:landscape} enable us to give predictions for the DC transport setup discussed above. Suppose we start from a configuration with a uniform pairing polarization and apply a voltage. As long as this configuration is locally stable, the current rises linearly with time. This proceeds until the current exceeds $J_c$. Then, the energy barrier for spontaneous dissipation vanishes, and a dissipation event would initiate. This process should repeat itself and the system would oscillate back and forth between a current close to the critical current $J\sim J_{c}$ and the relaxed state with $J\sim J_{c}-2J_0$; see Fig.~\ref{fig:sys}. Thus the observable mean current would be maintained close to 
\begin{equation}\label{eq:J_relaxed}
\bar{J}_c\defeq J_c-J_0=\frac{2\pi q(\kappa_m-\kappa_d)}{L_x}.
\end{equation}
At a finite voltage, the current may overshoot the critical value, since the uniform configuration is a saddle point of the free energy (despite being unstable). The escape process is triggered by thermals fluctuations and takes a finite time to occur. The larger the voltage, the more the current overshoots $J_c$, and hence the average current increases. Hence we anticipate
\begin{equation}\label{eq:pred}
\bar{J}\simeq\bar{J}_c+\mathcal{O}(V^\beta),
\end{equation}
with some exponent $\beta$. Note, that this estimate only holds for $\bar{J}_{c}\ge0$, since otherwise, $J_{c}$ is lesser than one winding.

\begin{figure}[t]
    \centering
    \includegraphics[width=1\linewidth]{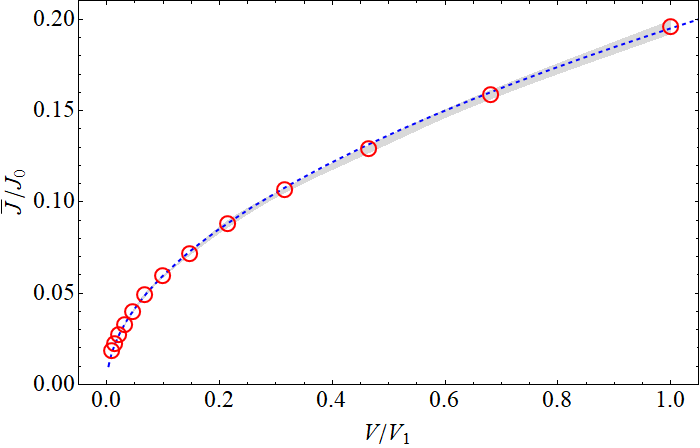}
    \caption{Numerics versus analytics: Current-voltage curve for $\kappa_m=\kappa_d$. Each circle depicts the average supercurrent from 18 instances of numerical simulations of Eq.~(\ref{eq:TDGLcross}); the gray area depict 95\% statistical confidence. The dashed line depicts the analytic results of Eq.~(\ref{eq:J_avg_exact}). Here, $V_1=\frac{\kappa_d\Gamma}{qL_x^2}$; exact details of the simulation are found in Appendix~\ref{app:numeric}. Remarkably, there are no fitting parameters.}
    \label{fig:exact-fit}
\end{figure}

In the following sections, we present both analytic and numeric results (see Figs.~\ref{fig:exact-fit} and \ref{fig:J_infinity}) showing these generic predictions are indeed realized in our specific TDGL model.

%%%%%%%%%%%%%%%%%%%%%%%%%%%%%%%%%%%
%%%%%%%%%%%%%%%%%%%%%%%%%%%%%%%%%%%

\subsubsection{Large dissipation limit}\label{sec:inf}
The simplest case to analyze is the case of $\frac{\Gamma\kappa_d}{L_x^2 qV}\to\infty$, where the dissipation rate is large compared with the Josephson frequency. In this case, the dependence of the current on time can be entirely found analytically (see Appendix~\ref{app:analytic}).

%%%%%%%%%%%%%%%%%%%%%%%%%%%%%%%%%%%
\begin{figure}[t]
    \centering
    \includegraphics[width=1\linewidth]{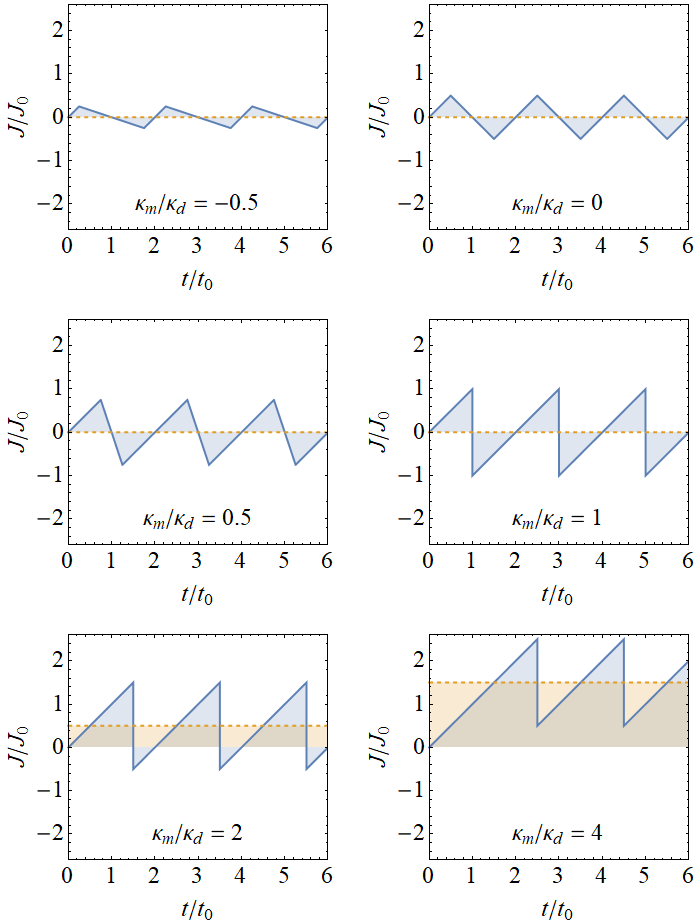}
    \caption{Supercurrent profiles $J(t)$ for infinite dissipation; see Sec.~\ref{sec:inf}. The average supercurrent $\bar{J}$ is depicted by the dashed line; the shaded areas depict the deviation from 0 supercurrent. Here, ${t_0}=\frac{2\pi}{qV}$ is the Josephson period.}
    \label{fig:J_infinity}
\end{figure}
%%%%%%%%%%%%%%%%%%%%%%%%%%%%%%%%%%%

In Fig.~\ref{fig:J_infinity} we show time-dependent supercurrent profiles $J(t)$ for various values of $\kappa_m/\kappa_d$, together with the corresponding values of the time-averaged supercurrent $\bar{J}$. For $\kappa_m\le\kappa_d$ the current profile is an asymmetric triangle wave pattern which averages out to 0, but for $\kappa_m >\kappa_d$ we have a shifted saw-tooth wave pattern which averages to a finite value. These patterns always exhibit a period of $2t_0=2(h/2eV)$, exactly matching our predictions for the doubling of the AC Josephson effect for small $B$; see Sec.~\ref{sec:results} and Eq.~\eqref{eq:period}. 

In this limit of large dissipation,
the time-averaged supercurrent is independent of voltage and temperature,
\begin{equation}\label{eq:Jinf}
\bar{J} =\begin{cases}
0 & \kappa_m\le\kappa_d\\
\frac{2\pi q(\kappa_m-\kappa_d)}{L_x} & \kappa_m>\kappa_d.
\end{cases}
\end{equation}
Due to the effectively infinite dissipation coefficient, the unwinding event begins immediately when the current reaches $J_{c}$, exactly matching our predictions of Eq.~\eqref{eq:Jc1}: there is no ``overshoot,'' even at a finite voltage. Therefore, the finite voltage correction discussed in Sec.~\ref{sec:pred} vanishes and we indeed get $\bar{J}\simeq\bar{J}_c$; cf. Eq.~\eqref{eq:J_relaxed}.
Note that the cusps in current vs. time are special to the infinite dissipation limit, and are smoothed out for finite dissipation; see Fig.~\ref{fig:JforV} in Appendix~\ref{app:analytic}.

\subsubsection{Finite dissipation}\label{sec:fin}

In the generic case of finite dissipation rate, 
we obtain a power-law relation between the time-averaged current $\bar{J}$ and the applied voltage.
We find that 
\begin{equation}
\label{eq:JV} \bar{J}-\bar{J}_c\propto q\sqrt{\tfrac{\kappa_d}{\Gamma}qV},
\end{equation}
cf.~Eq.~\eqref{eq:pred}.
This result holds for the supercurrent range of $\frac{1}{2}(\frac{\kappa_m}{\kappa_d}+3)-\frac{\bar{J}}{J_{0}}\gg\frac{L_x}{4\pi^2}\sqrt{\frac{qV}{\Gamma\kappa_d}}$, and $\bar{J}>|\bar{J}_c|$; see Eq.~\eqref{eq:Jcases} in Appendix~\ref{app:analytic}.
The dependence of $\bar{J}(V)$ on temperature is (sub-)logarithmic; the power-law dependence on $V$ in Eq.~(\ref{eq:JV}) may similarly be modified by logarithmic voltage corrections. Such logarithmic corrections are explicitly presented in the following subsection.

\subsubsection{Transition point}\label{sec:trans}
In the limit of infinite dissipation rate, Eq.~(\ref{eq:Jinf}) displays a non-analytic dependence of $\bar{J}$ on the ratio $\kappa_m/\kappa_d$, with a critical (transition) point at $\kappa_m = \kappa_d$.
For finite dissipation rates, the full equations for $\bar{J}$ are somewhat simplified at this transition point.
At this point we can obtain a complete solution; 
we find a characteristic voltage and temperature dependence of the average supercurrent,
\begin{equation}\label{eq:J_avg_exact}
\bar{J} =q\sqrt{\frac{\kappa_d qV}{\pi\Gamma}\ln\left(\frac{8L_y}{k_{B}T}\sqrt{\frac{\kappa_d qV}{\Gamma}}\right)}.
\end{equation}
This result is in excellent agreement with  numerical simulations of Eq.~(\ref{eq:TDGLcross}), as seen in Fig.~\ref{fig:exact-fit}; the details of the simulations are given in Appendix~\ref{app:numeric}.

Note that our result for the average current in Eq.~(\ref{eq:J_avg_exact}) na\"{i}vely diverges at $T\to0$. Although the current exceeds a level where there is no longer an energy barrier for the continuous unwinding process, at $T = 0$ there are no fluctuations to trigger decay within the classical model of Eq.~(\ref{eq:TDGLcross}).
However, at exponentially low temperatures where $\bar{J}\gg J_0$, spontaneous quantum decay process uncaptured by our analysis are expected to dominate the relaxation.

\section{Discussion and conclusions}\label{sec:discuss}
We end with a discussion of various points regarding the implications of our results in the context of polarized triplet SCs. In the following, we discuss the thermal phase diagram of two-dimensional spin-polarized SCs, possible mechanisms for relaxation of the magnetization (necessary for our topological mechanism for continuous current relaxation), the role of conventional supercurrent relaxation by vortex-antivortex dissociation, and estimated parameters for triplet superconductivity in graphene-based moir\'{e} materials. Finally, we point out some topics worthy of future studies.

\subsection{Thermodynamics of 2D spin-polarized superconductors} Throughout our discussion we have assumed that, at equilibrium, our system is essentially long-range ordered, i.e., $\langle \dd \rangle\ne 0$. This is, of course, never strictly true in two spatial dimensions at any non-zero temperature. In the absence of a Zeeman field and any other breaking of spin rotational symmetry, the system is always in a disordered phase at all $T>0$~\cite{Mukerjee2006}, in accordance with the Mermin-Wagner theorem. However, the spin correlation length grows rapidly with decreasing temperature, $\xi\sim e^{2\pi(\kappa_m + \kappa_d)/(k_B T)}$; hence, at sufficiently low temperatures, $\xi \gg \max(L_x,L_y)$ and our treatment should apply. If $\kappa_m$ is larger than $\kappa_d$, then there is a finite temperature crossover for finite systems that resembles a BKT 
transition. 
In the presence of a Zeeman field that pins the direction of the pairing polarization, the system undergoes a genuine finite-temperature BKT transition. 

We note in passing that, in contrast to a fully polarized triplet SC [phase (ii) in Sec.~\ref{sec:sym}], a partially polarized phase [phase (iii) in Sec.~\ref{sec:sym}] undergoes a true BKT transition at finite temperature even for $\vec{B}=0$, at which $\Delta_{4e}$ becomes quasi-long range ordered~\cite{Mukerjee2006}. This phase is interesting in its own right, as it supports half quantum vortices (carrying flux $h/4e$). We leave a more detailed discussion of this phase and its physical implications for future work.  

\subsection{Mechanism for magnetization relaxation}\label{sec:mag}

The continuous supercurrent relaxation process discussed in this work requires a mechanism
for the pairing polarization to dynamically change in time (both locally and globally). Since the pairing polarization is pinned to the physical spin polarization, this requires a mechanism for the system to exchange spin angular momentum with its environment. In our time-dependent model (Sec.~\ref{sec:tdgl}), we assumed the existence of a spin bath on empirical grounds. In practice, the maximum rate at which the continuous relaxation process can proceed, $1/\tau_s$, depends on the physical mechanism of spin relaxation~\footnote{Within the time-dependent model discussed in Sec.~\ref{sec:tdgl}, $1/\tau_s$ corresponds to the dissipation rate, Eq.~\eqref{eq:diss} in Appendix~\ref{app:analytic}.}. The system's spin can relax either through spin-orbit coupling (which also introduces anisotropic terms in spin space into the free energy), a spin bath due to internal or substrate impurities or nuclear spins, or through the boundaries, in cases where the system is coupled to metallic leads. 

In the context of TBG and related materials, spin-orbit coupling is expected to be weak, and nuclear spins are rare (unless \textsuperscript{13}C impurities are introduced intentionally). In the absence of other magnetic impurities, the most dominant source of spin relaxation is likely by spin transport to the metallic leads. This mechanism is absent in our simple dynamical model of Sec.~{\ref{sec:tdgl}} and we leave its detailed study to follow up work.
An appropriate description of this mechanism might be the use of either the solitons of the degeneracy space~\cite{Tjon1977Solitons,Lamacraft2017Persistent} or the Goldstone modes of the order parameter in order to carry the magnetic texture to the leads. 

Here, we make a rough estimate of the spin relaxation time due to this mechanism. For simplicity's sake, we assume that there is no applied Zeeman field and set $B=0$. Given that the spin stiffness in our model is $\kappa_m+\kappa_d$ [see Eq.~\eqref{eq:Fd_app}], we expect on dimensional grounds 
that $1/\tau_s \sim \frac{(\kappa_m+\kappa_d) \xi_m^2}{ L_x^2}$, where $\xi_m$ is a microscopic ``magnetic coherence length'' (essentially, $\mu_B/\xi_m^2$ is the magnetization density) and $L_x$ is the distance between the leads.
This estimate for $\tau_s$ can be understood as the time it takes for a quadratically dispersing magnon with wavevector $|\vec{k}|\sim 1/L_x$ to travel across the system.

For a sufficiently small applied voltage such that $eV \ll 4\pi\hbar/\tau_s$, the continuous unwinding process can take place within one (doubled) Josephson period, with spin carried in and out of the system through its connection to the leads. For larger voltages, the continuous unwinding process is limited by the rate of spin relaxation. The study of this case goes beyond our present analysis. A possible scenario, in this case, is that the phase accumulates more than two windings between the continuous unwinding events, decreasing the effective Josephson frequency below $\tilde{\omega}_J = eV/\hbar$.

To get a rough estimate of $1/\tau_s$ due to spin dissipation at the boundaries in graphene-based moir\'{e} materials, we assume that the spin stiffness $\kappa_m+\kappa_d$ is of the order of $0.1-1$ meV and $\xi_m$ is of the order of a few times the moir\'{e} lattice spacing $a\approx 10$ nm. For example, suppose that $\kappa_m+\kappa_d=1$~meV and $\xi_m = 5a$. Then, for a system of size $L_x = 1\,\mu$m, the above considerations give that the crossover voltage is $V\approx 4\pi\hbar /(e \tau_s)  \approx 30\mathrm{\,\mu V}$.

\subsection{Phase unwinding due to vortex motion}\label{sec:vortex}

In our analysis of the supercurrent relaxation through continuous unwinding, we have neglected the ordinary mechanism of vortex-antivortex dissociation and motion of free vortices. 
In two spatial dimensions, this mechanism gives rise to the celebrated nonlinear $I$-$V$ characteristics~\cite{halperin1979resistive,Ambegaokar1980Dynamics,Epstein1981Vortex,Kadin1983Renormalization,newrock2000two} associated with the BKT transition: $V\propto I^{\alpha}$ with $\alpha(T)=\frac{\pi\kappa_s(T)}{k_B T}+1$, where the phase stiffness $\kappa_s$ satisfies $\kappa_s(T_{\mathrm{BKT}})=\frac{2}{\pi}k_B T_{\mathrm{BKT}}$ such that $\alpha(T_{\mathrm{BKT}})=3$. In our model, $\kappa_d=\frac{1}{2}\kappa_s$, this provides us with the estimates in Eq.~\eqref{eq:main}.

In order to determine which mechanism dominates the supercurrent relaxation, we need to compare the rate of phase unwinding (i.e., the voltage) generated by vortex motion to the rate of the continuous unwinding process. The voltage due to vortex motion is estimated as~\cite{halperin1979resistive}
\begin{equation}\label{eq:Vc}
V_{\text{vm}}\approx 2 \rho_n  J L_x \left(\alpha(T)-3\right)\left[\max\left(\frac{\xi}{L},\frac{J}{J_{c,0}}\right)\right]^{\alpha(T)-1}.
\end{equation}
Here, $\rho_n$ is the normal state resistivity, $J$ is the current density,  $\xi$ is the coherence length, $L= \min(L_x,L_y)$, and $J_{c,0}\sim\frac{e\,k_B T_{\mathrm{BKT}}}{\hbar \xi}$ is microscopic critical current density. 
The continuous unwinding mechanism occurs when $J\gtrsim J_c$, whereby it does not involve passing through an energy barrier. Hence, the continuous unwinding mechanism is dominant when $V\gg V_{\text{vm}}(J_c)$. 
Therefore, as long as $J_c<J_{c,0}$, one sees that $V_{\text{vm}}(J_c)  \rightarrow 0$ at low temperature, since the vortex-antivortex unbinding process is thermally activated. 

\subsection{Twisted double-bilayer graphene parameters}

As discussed in the Introduction, recent experiments in TDBG with a perpendicular electric field revealed a ferromagnetic ground state at half-filling of the moir\'{e} lattice~\cite{liu2019spin,cao2020tunable}. Upon changing the density away from half-filling, the resistance drops dramatically, possibly due to superconductivity~\cite{shen2020correlated,liu2019spin}. Most strikingly, the temperature at which the resistivity drops increases linearly as a function of an in-plane magnetic field for small fields, a signature of \emph{triplet} superconductivity~\cite{liu2019spin,lee2019theory,scheurer2019pairing,Samajdar2020microscopic,Hsu2020topological,Wu2019Identification}. Note, however, that an interpretation of these observations in terms of a non-superconducting state has also been proposed~\cite{he2020tunable}.

Specifically for this case of TDBG, we may estimate the values of the various parameters in our model using the experimental data of Ref.~\onlinecite{liu2019spin}. 
The sample dimensions in these experiments were of the order of a few $\mu$m, 
with a carrier density of $n\approx 2\times 10^{12}\mathrm{\,cm^{-2}}$. As discussed in Sec.~\ref{sec:rev}, we focus on dynamics far below the critical temperature $T_{\mathrm{BKT}}\approx 3.5\mathrm{\,K}$.
For an order of magnitude estimate, we set $\kappa_d\sim\frac{1}{2}\kappa_s(T_{\mathrm{BKT}})=\frac{1}{\pi}k_B T_{\mathrm{BKT}}\approx 0.1\mathrm{\,meV}$.
We furthermore estimate the magnetic moment density as $\mu\approx n\mu_B$.

Using these estimated parameter values, we assess the two regimes of critical current behavior in Eq.~\eqref{eq:Jc1}.
We find $J_c\approx 300 \frac{\mathrm{nA}}{\mathrm{\mu m}}\times\sqrt{\frac{B}{\mathrm{mT}}}$ for large applied Zeeman field, $B$, and $I_c\approx 500\,\mathrm{nA}\times \frac{L_y}{L_x}$ for small $B$. The crossover between the two regimes occurs at $B\approx 3\mathrm{\,mT}\times\frac{\mathrm{\mu m}^2}{L_x^2}$.

\subsection{Conclusions and future directions}

In this paper, we have studied the transport properties of fully spin-polarized triplet SCs. Interestingly, we found that the topology of the order parameter degeneracy space can have dramatic consequences, given that a mechanism for spin dissipation is available. Basically, two windings of the order parameter are topologically equivalent to zero windings, and hence a supercurrent can decay via a smooth deformation of the order parameter that involves a transient inhomogeneous spin texture but does not involve creating topological defects (such as vortices). 
We expect that this mechanism of supercurrent relaxation 
may become dominant at sufficiently low temperatures, where vortex configurations are suppressed. 

A direct observable manifestation of the continuous supercurrent relaxation mechanism is that a sample subjected to a DC voltage exhibits an oscillatory current with a fundamental frequency which is a fraction of the Josephson frequency. Furthermore, an applied Zeeman field suppresses the continuous supercurrent decay mechanism, since it pins the direction of the system's magnetization.

A possible extension of our analysis is the study of the equilibrium properties of the system in the presence of a perpendicular magnetic field. At high applied field one would find the standard vortex lattice groundstate. However, at low magnetic fields, one would suspect a stabilization of a vortex-free spin texture; this would be reminiscent of Fig.~\ref{fig:traj} in space rather than in time. In addition, similar physics to that discussed here might arise in superconducting states that emerge out of more exotic spin-valley ferromagnets, such as the skyrmion SC discussed in Ref.~\cite{Khalaf2020skyrmions}.
We leave these directions to future work.

\section*{Acknowledgments}
We are grateful for illuminating discussions with A.~Auerbach, J.~Ruhman, O.~Golan, S.~Kivelson, D.~Podolsky, T.~Senthil, and A.~Vishwanath. EB and EC were supported by the European Research Council (ERC) under grant HQMAT (Grant Agreement No.~817799) and the US-Israel Binational Science Foundation (BSF). EB and MR acknowledge support from CRC 183 of the Deutsche Forschungsgemeinschaft. MR gratefully acknowledges the support of the European Research Council (ERC) under the European Union Horizon 2020 Research and Innovation Programme (Grant Agreement No.~678862), and the Villum Foundation.

\appendix
\section{Fluctuation-dissipation theorem}\label{app:FDT}
In this Appendix, we show that the Langevin equation [Eq.~\eqref{eq:TDGLindex}] satisfies the fluctuation-dissipation theorem (FDT) at equilibrium. Recall that a sufficient condition for a stationary stochastic process to satisfy the FDT is that the Gibbs distribution, $W\propto e^{-F/{k_B T}}$, is time-independent under the stochastic process~
\cite{FDT2020}. This means that the Gibbs distribution is a  
solution of the corresponding Fokker-Planck equation. 

We consider a set of dynamical variables $X_I$ which satisfy general Langevin equations of the form (using the Stratonovich convention~\cite{risken1996fokker})
\begin{equation}
    \frac{\partial X_{I}}{\partial t}=h_{I}(\{X\})+g_{I\mu}(\{X\})\zeta_{\mu}(t),
\end{equation}
where repeated indices are summed over; $h_I$, $g_{I\mu}$ are differentiable functions, and $\zeta_{\mu}(t)$ are independent Gaussian noise sources with zero mean,  satisfying $\langle \zeta_{\mu}(t)\zeta_{\nu}(t)\rangle = k_B T\Gamma  \delta_{\mu\nu}\delta(t-t')$. The corresponding Fokker-Planck equation that describes the evolution of the distribution function $W(\{X\},t)$ is~\cite{risken1996fokker}:
\begin{multline}
\frac{\partial W}{\partial t}(\{X\},t)=-\frac{\partial}{\partial X_{I}}\left[h_{I}(\{X\})W\right] \\
+k_B T\Gamma\frac{\partial}{\partial X_{I}}\left\{ g_{I\mu}(\{X\})\frac{\partial}{\partial X_{J}}\left[g_{J\mu}(\{X\})W\right]\right\}. 
\end{multline}
Applying this rule to our Langevin equation [Eq.~\eqref{eq:TDGLindex}] with $X_I\mapsto d_i^{(\alpha)}(\rr)$, we obtain:
\begin{align}\label{eq:FP}
\frac{\partial W}{\partial t}(\{d\},t)    & =\nonumber\\ 
- & \frac{\partial}{\partial d_{i}^{(\alpha)}}\left\{ \left[\varepsilon^{\alpha\beta}\frac{\delta F}{\delta d_{i}^{(\beta)}}-\Gamma g_{ik}^{\alpha}g_{jk}^{\beta}\frac{\delta F}{\delta d_{j}^{(\beta)}}\right]W\right\}\nonumber\\ 
+ & k_{B}T\Gamma\frac{\partial}{\partial d_{i}^{(\alpha)}}\left\{ g_{ik}^{\alpha}\frac{\partial}{\partial d_{j}^{(\beta)}}\left[g_{jk}^{\beta}\,W\right]\right\} .
\end{align}
Substituting $W = e^{-F[\{d\}]/k_BT}/Z$ and using the fact that for our choice $g^{\alpha}_{ik} = \varepsilon_{ijk} d^{(\alpha)}_j$ [see discussion below Eq.~(\ref{eq:TDGLindex})]:
\begin{equation}
\frac{\partial g_{jk}^{\beta}}{\partial d_{j}^{(\beta)}}=0,    
\end{equation}
we find that the Gibbs distribution is a stationary solution of Eq.~(\ref{eq:FP}). Hence the FDT is satisfied.\\

\section{Analytic solutions to the TDGL equations}\label{app:analytic}

In this appendix, we present the details of the analytic solution to Eq.~(\ref{eq:TDGLcross}).
\subsection{Preliminaries}
We focus on the case of large dissipation $\Gamma\gg d_0^{-2}$.
In this case there are three
frequency scales in the system. The 
inverse of the Josephson period,
\begin{equation}
\frac{1}{t_0}=\frac{qEL_x}{2\pi},
\end{equation}
the dissipation rate,
\begin{equation}\label{eq:diss}
\gamma=\frac{4\pi^{2}\kappa_d\Gamma}{L_x^2},
\end{equation}
and the fluctuation rate,
\begin{equation}
\T=\frac{2 k_B T \Gamma}{L_x L_y}.
\end{equation}
Whenever possible we shall express our equations and results in terms of these quantities.\\

\subsection{The equations}\label{sec:eq}
We explicitly take ${\dd\cdot{\dd}}=0$, characteristic of the fully spin-polarized phase, as discussed in Sec.~\ref{sec:sym}; this condition is conserved by the equations of motion.
We furthermore take the London limit of  ${\dd\cdot{\dd}^{\ast}}=2d_{0}^{2}$. These conditions imply $U(\dd)=const$ and $\nabla\cdot \J=-\dot{\rho}=0$. This allows one to rewrite the free energy as
\begin{equation}\label{eq:Fd_app}
F[\dd] =\int d^{2}r\left\{\frac{\kappa_d+\kappa_m}{2d_{0}^{2}}\left|\left(\nabla-iq\A\right)\dd\right|^{2}
-\frac{\kappa_m}{4\kappa_d^2}\J^2\right\},
\end{equation}
and to explicitly write the equations of motion [Eq.~(\ref{eq:TDGLcross})],
\begin{widetext}
\begin{align}
\frac{\partial \dd}{\partial t} = \ & \Gamma\frac{\kappa_d+\kappa_m}{2}\bigg[2\left(\nabla^{2}-2iq\A\cdot\nabla\right)\dd-\frac{1}{d_{0}^{2}}\left({\dd}\cdot\left(\nabla^{2}-2iq\A\cdot\nabla\right)\dd\right){\dd}^{\ast}-\frac{1}{d_{0}^{2}}\left({\dd}\cdot\left(\nabla^{2}+2iq\A\cdot\nabla\right)\dd^{\ast}\right){\dd}\bigg]\nonumber \\
& -i\Gamma\frac{\kappa_m}{2\kappa_d}\left[2\nabla\dd+\frac{1}{d_{0}^{2}}\left({\dd}\cdot\nabla{\dd}^{\ast}\right){\dd}\right]\J+{\dd}\times\bm{\zeta}.\label{eq:transport-PDE}
\end{align}
\end{widetext}
This nonlinear partial
differential equation is best solved by the method of guessing the
solution. We use an ansatz inspired by the unwinding trajectory of Sec.~\ref{sec:landscape},
\begin{align}\label{eq:ansatz}
u & =u_{0}e^{ i\sigma_{3}\left(\pi\Delta\frac{x}{L_x}+\frac{\phi_{\Delta}}{2}\right)} e^{ i\sigma_{1}\left(\frac{1}{2}f(t)+\frac{\pi}{4}\right)} e^{ i\sigma_{3}\left(-\pi n\frac{x}{L_x}+\frac{\phi_{n}}{2}\right)} .
\end{align}
Here, $n$ and $\Delta$ are integers, while $\phi_n$ and $\phi_\Delta$ are some arbitrary phases, and $u_0$ is an arbitrary $SU(2)$ matrix.

The current is given by
\begin{equation}\label{eq:f2J}
J(t) =J_{0}\left[\tfrac{t}{t_{0}}-\left(n+\Delta\sin f(t)\right)\right].
\end{equation}

This trajectory connects two families of constant configurations; $f_-=-\frac{\pi}{2}+2\pi\ell$ with $n-\Delta$ windings, and $f_+=\frac{\pi}{2}+2\pi\ell$ with $n+\Delta$ windings, where $\ell\in\mathbb{Z}$. Note that these configurations have uniform pairing-polarization and hence the winding number is unambiguously defined. Both configurations have supercurrent profiles growing linearly with time,
$J_\pm(t)=J_{0}[\frac{t}{t_{0}}-(n\pm\Delta)]$.

The dynamics of the system consist of a series of unwinding events with different values of parameters $n,\Delta,\phi_{n,\Delta},u_0$. The pairing-polarization and supercurrent at the end of an unwinding event determine $u_0$ and $n-\Delta$ of the next event. We assume the system is initialized with $J(t=0)=0$ such that $n=\Delta$ for the first unwinding event.

\subsection{Zero temperature}\label{sec:T0}
First, we analyze the case of $T=0$, whereby our ansatz [Eq.~\eqref{eq:ansatz}] reduces the equations of motion to an ordinary differential equation (ODE),
\begin{equation}
\dot{f}(t) =\gamma\Delta\left\{ 2\left(\tfrac{t}{t_{0}}-n\right)-\Delta\left(1-\tfrac{\kappa_m}{\kappa_d}\right)\sin f(t)\right\} \cos f(t).
\end{equation}
There are two families of fixed-point solutions to this equation, corresponding to the constant configurations discussed above, i.e., $f_{\pm}=\pm\frac{\pi}{2}+2\pi\ell$, where $\ell\in\mathbb{Z}$. However,
when $t\to\infty$ only $f_{+}$ are stable fixed-points under small perturbations
while $f_{-}$ are unstable;
when $t\to-\infty$ only $f_{-}$ are stable fixed-points while $f_{+}$ are unstable.
Therefore, a solution to this equation describes a trajectory from $f_{-}$ to $f_{+}$. This trajectory describes relaxation of $2\Delta$ windings, and the fundamental 2-windings relaxation trajectory is attained for $\Delta=1$.

At an intermediate time, $t_\mathrm{init}$, when $J(t_\mathrm{init})=J_{c}$ [see Eq.~\eqref{eq:Jc1}], the stable fixed point $f_-$ at $t\to-\infty$ becomes a saddle-point. Any small deviation would thus send the system on a trajectory towards the stable fixed point $f_+$ at $t\to\infty$, which itself ceases being a saddle-point only at some other intermediate time ($t_\mathrm{fin}$ discussed below).

However, at $T=0$, there is no source for small deviations and the system would thus remain frozen at $f_-$. Moreover, even given some initial perturbation, it could only trigger the first unwinding event and the system would eventually remain asymptotically close to $f_+$ with nothing to trigger the next unwinding event. Therefore it is crucial to study the effects of finite temperature.

\subsection{Effective model for finite temperatures}
At finite temperatures, $T>0$, the thermal fluctuations would nudge the system from its saddle-point and initiate an unwinding trajectory. The system may spontaneously choose any random fluctuation direction $\mathbf{v}(x)$ within the manifold of configurations escaping the saddle-point. Without loss of generality, we pick such a direction $\mathbf{v}(x)$ compatible with our ansatz [Eq.~\eqref{eq:ansatz}],
\begin{equation}
\mathbf{v}(x)=\begin{pmatrix}
\cos\left(2\pi\Delta\frac{x}{L_x}+\phi_{\Delta}\right)\\
-\sin\left(2\pi\Delta\frac{x}{L_x}+\phi_{\Delta}\right)\\
0
\end{pmatrix}.
\end{equation}
This is supported by numerical simulations of Eq.~(\ref{eq:transport-PDE}) for finite temperatures.

This choice of direction enables us to drastically simplify the functional vector noise term in Eq.~(\ref{eq:TDGLindex}) and replace it with an effective uniform scalar noise term, $g^{\alpha}_{ij}(\dd)\zeta_j(\rr,t) \mapsto \varepsilon_{ijk} d^{\alpha}_j v_k(x)\zeta(t)$, with
\begin{equation}
\left\langle \zeta(t)\zeta(t')\right\rangle =  2k_{B}T\Gamma_{\Delta}\delta(t-t').
\end{equation}
The corresponding equations of motion are modified accordingly (see Appendix~\ref{app:FDT}),
\begin{widetext}
\begin{align}
\frac{\partial \dd(\rr,t)}{\partial t}= \ & \Gamma_{\Delta}\frac{\kappa_d+\kappa_m}{2d_{0}^{2}}\left\{\int d^{2}r'\left[ \left({\dd}^{\ast}\times\mathbf{v}\right)\cdot\left(\nabla^{2}-2iq\A\cdot\nabla\right)\dd+\left({\dd}\times\mathbf{v}\right)\cdot\left(\nabla^{2}+2iq\A\cdot\nabla\right){\dd}^{\ast}\right]_{\rr'} \right\}\left({\dd}(\rr,t)\times\mathbf{v}(x)\right)\nonumber \\
& -i\Gamma_{\Delta}\frac{\kappa_m}{2d_{0}^{2}\kappa_d}\left\{\int d^{2}r'\left[ \left(\left({\dd}^{\ast}\times\mathbf{v}\right)\cdot\nabla\dd-\left({\dd}\times\mathbf{v}\right)\cdot\nabla{\dd}^{\ast}\right)\J\right]_{\rr'} \right\}\left[{\dd}(\rr,t)\times\mathbf{v}(x)\right]+{\dd}(\rr,t)\times\mathbf{v}(x)\zeta(t).
\end{align}
\end{widetext}
As we show in Appendix~\ref{sec:rate}, consistency requires $\Gamma_{\Delta} =\Gamma/L_x L_y$.

Remarkably, this effective noise term propagates well into our ansatz [Eq.~(\ref{eq:ansatz})],
\begin{align}\label{eq:fzeta}
\dot{f}(t) = \ &\gamma\Delta\left\{ 2\left(\tfrac{t}{t_{0}}-n\right)-\Delta\left(1-\tfrac{\kappa_m}{\kappa_d}\right)\sin f(t)\right\} \cos f(t) \nonumber\\
&+\zeta(t),
\end{align}
thus reducing the equations of motion to a stochastic ODE.
As seen in Fig.~\ref{fig:exact-fit-dev} in Appendix~\ref{app:numeric}, this effective model exquisitely captures the numerical detail of Eq.~(\ref{eq:TDGLcross}).

\subsubsection{Evaluation of the diffusion rate}\label{sec:rate}

Given our original 2+1 dimensional vector model, 
\begin{equation}
\avg{\zeta_i(\rr,t)\zeta_j(\rr',t')} =2k_{B}T\Gamma\delta_{ij}\delta(\rr-\rr')\delta(t-t'),
\end{equation}
and an effective 0+1 dimensional scalar model fluctuating in the $\mathbf{v}(x)$ direction,
\begin{equation}
\left\langle \zeta(t)\zeta(t')\right\rangle =2k_{B}T\Gamma_{\Delta}\delta(t-t'),
\end{equation}
we wish to find the value of $\Gamma_\Delta$ such that the effective scalar model best approximates the original vector model.

Hence, we project $\bm{\zeta}(\rr,t)$ onto $\mathbf{v}(x)$, i.e.,
\begin{equation}
\zeta(t)  =\frac{\int_{0}^{L_y}dy\int_{0}^{L_x}dx\,\mathbf{v}(x)\cdot\bm{\zeta}(\rr,t)}{L_y \int_{0}^{L_x}dx\,\mathbf{v}(x)\cdot\mathbf{v}(x)},
\end{equation}
such that
\begin{align}
\avg{\zeta(t)\zeta(t')}   &=\frac{\iint d^2r \iint d^2r'\mathbf{v}(x)\cdot\left\langle \bm{\zeta}(\rr,t)\bm{\zeta}(\rr',t')\right\rangle \cdot\mathbf{v}(x')}{L_x^{2}L_y^2} \nonumber\\
&=\frac{1}{L_x L_y}2k_{B}T\Gamma\delta(t-t'),
\end{align}
and thus find $\Gamma_{\Delta} =\Gamma/L_xL_y$.

\subsection{Solutions of the effective model}
We are now finally in a position to derive the results of Sec.~\ref{sec:sol}.
We solve Eq.~(\ref{eq:fzeta}) and use Eq.~(\ref{eq:f2J}) and Eq.~(\ref{eq:Javg}) to obtain the average supercurrent $\bar{J}$ for various ratios of the rates $t_0^{-1},\gamma,\T$ and various ratios of the stiffnesses $\kappa_d,\kappa_m$.

\subsubsection{Infinite dissipation}\label{sec:app_inf}
The simplest solution is when $t_{0}\gamma\to\infty$.
In such cases
the solution is 
\begin{align}
&\kappa_m<\kappa_d: && \sin f(t)=\begin{cases}
-1 & t<t_{\mathrm{init}},\\
-1+2\frac{t-t_{\mathrm{init}}}{t_{\mathrm{fin}}-t_{\mathrm{init}}} & t_{\mathrm{init}}\le t\le t_{\mathrm{fin}},\\
+1 & t>t_{\mathrm{fin}},
\end{cases} \nonumber\\
&\kappa_m\geq\kappa_d: && \sin f(t)=\begin{cases}
-1 & t<t_{\mathrm{init}},\\
+1 & t>t_{\mathrm{init}},
\end{cases}
\end{align}
where we have defined
\begin{align}
t_{\mathrm{init}} & = t_{0}\left(n-\frac{\Delta}{2}\left(1-\frac{\kappa_m}{\kappa_d}\right)\right),\nonumber\\
t_{\mathrm{fin}} & = t_{0}\left(n+\frac{\Delta}{2}\left(1-\frac{\kappa_m}{\kappa_d}\right)\right).
\end{align}

By matching the initial and final conditions discussed in Appendix~\ref{sec:eq} we find that all unwinding events have $\Delta=1$ (corresponding to the fundamental 2 windings relaxation event), and that $n=1,3,5,\ldots$ . For $\kappa_m\le\kappa_d$ the current profile is an asymmetric triangle wave pattern which averages out to 0, but for $\kappa_m>\kappa_d$ we have a shifted saw-tooth wave pattern for $J(t)$ and 
$\bar{J}=J_{0}\frac{[t_{\mathrm{init}}]_{n=1}-t_{0}}{t_{0}}=J_{0}\frac{1}{2}\left(\frac{\kappa_m}{\kappa_d}-1\right)$. 
We thus conclude:
\begin{equation}
\bar{J} =\begin{cases}
0 & \kappa_m\le\kappa_d,\\
J_{0}\frac{1}{2}\left(\frac{\kappa_m}{\kappa_d}-1\right) & \kappa_m>\kappa_d.
\end{cases}
\end{equation}
This is precisely Eq.~\eqref{eq:Jinf} of the main text, with the corresponding behavior plotted in Fig.~\ref{fig:J_infinity}.

\subsubsection{Finite dissipation}\label{sec:finite_dissip}

For finite $t_{0}\gamma$, we linearize Eq.~\eqref{eq:fzeta} around $f(t)=-\frac{\pi}{2}+h(t)$ with small $h(t)$ such that 
\begin{equation}
\dot{h}(t) \simeq2\gamma\Delta\frac{t-t_{\mathrm{init}}}{t_{0}}h(t)+\zeta(t).
\end{equation}
This is solved by
\begin{align}
h(t)  = \ &e^{\frac{\gamma\Delta}{t_{0}}(t-t_{\mathrm{init}})^{2}}\int_{-\infty}^{t-t_{\mathrm{init}}}dt'e^{-\frac{\gamma\Delta}{t_{0}}t'^{2}}\zeta(t'),\\
\avg{h(t)^{2}}  = \ &e^{2\frac{\gamma\Delta}{t_{0}}(t-t_{\mathrm{init}})^{2}} \nonumber\\
&\times\int\limits_{-\infty}^{t-t_{\mathrm{init}}}dt'\int\limits_{-\infty}^{t-t_{\mathrm{init}}}dt'' 
e^{-\frac{\gamma\Delta}{t_{0}}(t'^2+t''^2)}\T\delta(t'-t'')\nonumber \\
= \ & \T e^{2\frac{\gamma\Delta}{t_{0}}(t-t_{\mathrm{init}})^{2}}\int_{-\infty}^{t-t_{\mathrm{init}}}dt'e^{-2\frac{\gamma\Delta}{t_{0}}t'^{2}},\\
h_{\mathrm{init}}^{2}  = \ &\avg{h(t_{\mathrm{init}})^{2}} =\T\frac{1}{2}\sqrt{\frac{\pi}{2}}\sqrt{\frac{t_{0}}{\gamma\Delta}}.\label{eq:hinit}
\end{align}
Here we have approximated $h(t\to-\infty)\to0$. This approximation requires contributions from previous unwinding events, which end at time $t_{\mathrm{prev}}$, 
to decay before the initiation of a new event:
\begin{equation}\label{eq:forget}
1\gg\frac{\avg{h(t_{\mathrm{prev}})^{2}}}{\avg{h(t_{\mathrm{init}})^{2}}}\simeq\frac{\sqrt{\frac{t_0}{2\pi\gamma\Delta}}}{t_{\mathrm{init}}-t_{\mathrm{prev}}}.
\end{equation}
We evaluate $t_\mathrm{prev}$ and check our solutions for this condition at Appendix~\ref{sec:app_generic}.

Moreover, the linearization requires
\begin{equation}\label{eq:lowtemp}
h_\mathrm{init}\ll\pi,\quad\text{where}\quad h_\mathrm{init}\sim\T^{1/2}(t_0/\gamma)^{1/4}.
\end{equation}
This is the low-temperature condition discussed in Sec.~\ref{sec:sol}. It ensures there are well separated unwinding events as the temperature is not so high that the system escapes the stable fixed-points at any time by thermal fluctuations.

\subsubsection{Exactly solvable point}

The nonlinear ordinary differential equation Eq.~(\ref{eq:fzeta}) becomes an exact differential equation at $\kappa_m=\kappa_d$.

First, we set $\Delta=1$ as in the $t_{0}\gamma\to\infty$ case (see discussion in Appendix~\ref{sec:app_generic}).
Next, we use the linearized equation to find the initial displacement [Eq.~\eqref{eq:hinit}].
Then, starting from $f(t_{\mathrm{init}})\simeq-\frac{\pi}{2}+h_{\mathrm{init}}$, we propagate in time, neglecting the effects of the noise term,
\begin{align}
\dot{f}(t)  \simeq \ & 2\gamma\frac{t-t_{\mathrm{init}}}{t_{0}}\cos f(t),\\
f(t>t_{\mathrm{init}})  = \ & 2\arctan\tanh\bigg(\frac{\gamma}{2t_{0}}(t-t_{\mathrm{init}})^{2} \nonumber\\
&-\mathrm{arctanh}\tan\frac{\frac{\pi}{2}-h_{\mathrm{init}}}{2}\bigg).
\end{align}
We may hence easily evaluate the average supercurrent,
\begin{align}
\bar{J} & \simeq\frac{1}{2t_{0}}\int_{t_{0}}^{\infty}dt\left\{ J(t)-J_{0}\left(\frac{t}{t_{0}}-2\right)\right\} \nonumber \\
& =\frac{J_{0}}{2t_{0}}\int_{-\infty}^{\infty}dt'\frac{1}{1+e^{\frac{2\gamma}{t_{0}}t'^{2}}\tan^{2}\frac{h_{\mathrm{init}}}{2}}\nonumber \\
& =-J_{0}\sqrt{\frac{\pi}{8\gamma t_{0}}}\mathrm{Li}_{\frac{1}{2}}\left(-\cot^{2}\frac{h_{\mathrm{init}}}{2}\right).
\end{align}
Here, $\mathrm{Li}_{s}(z)$ is the polylogarithm function. As we have taken $h_{\mathrm{init}}\ll\pi$ we may use $\mathrm{Li}_{s}(-z)\simeq-\frac{1}{\Gamma(s+1)}\ln z$
and get
\begin{equation}\label{eq:J_exact}
\bar{J} \simeq J_{0}\sqrt{\frac{1}{2\gamma t_{0}}\ln\frac{8}{\T\sqrt{\frac{\pi}{2}}\sqrt{\frac{t_{0}}{\gamma}}}}.
\end{equation}
This is Eq.~(\ref{eq:J_avg_exact}) of the main text.

\subsubsection{Generic case}\label{sec:app_generic}

At large but finite dissipation we wish to find the approximate escape times $t_{\ast}>t_{\mathrm{init}}$ when the linearization breaks down and $h(t_\ast)\sim\frac{\pi}{2}$ is no longer small.

%%%%%%%%%%%%%%%%%%%%%%%%%%%%%%%
\begin{figure}[t]
    \centering
    \includegraphics[width=1\linewidth]{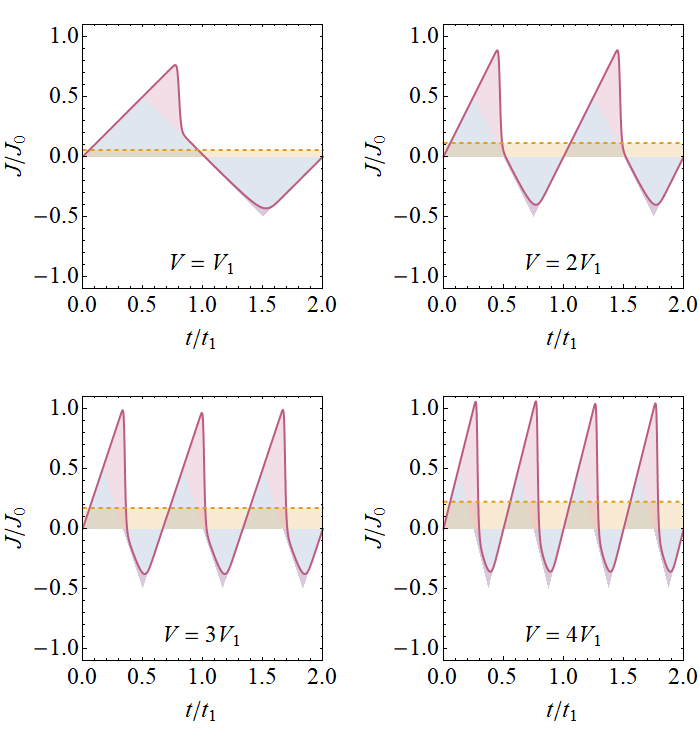}
    \caption{Supercurrent profiles $J(t)$ for various voltages; see Sec.~\ref{sec:fin}. The average supercurrent $\bar{J}$ is depicted by the dashed line; the shaded areas depict the deviation from the infinite dissipation limit; see Fig.~\ref{fig:J_infinity}. Here, ${t_1}=\frac{2\pi}{qV_1}$; exact details of the simulation are found in Appendix~\ref{app:numeric}.}
    \label{fig:JforV}
\end{figure}
%%%%%%%%%%%%%%%%%%%%%%%%%%%%%%%

We focus on 
\begin{equation}\label{eq:cond1}
t_{\ast}-t_{\mathrm{init}}\gg\sqrt{\frac{t_{0}}{2\gamma}},
\end{equation}
where one finds
\begin{align}
h_{\ast}^{2} & =\avg{h(t_{\ast})^{2}} \simeq \T\sqrt{\frac{\pi}{2}}\sqrt{\frac{t_{0}}{\gamma\Delta}}e^{2\frac{\gamma\Delta}{t_{0}}(t_{\ast}-t_{\mathrm{init}})^{2}},\nonumber \\
t_{\ast} & \simeq t_{\mathrm{init}}+\sqrt{\frac{t_{0}}{2\gamma\Delta}\ln\frac{h_{\ast}^{2}}{\T\sqrt{\frac{\pi}{2}}\sqrt{\frac{t_{0}}{\gamma\Delta}}}}.\label{eq:t_ast}
\end{align}
The first unwinding event satisfies $n=\Delta$. 
The following cascade of unwinding events is very complicated. 
Although this case looks interesting, we suspend its analysis
to further research. We thus stick for now with the assumption that
$\Delta=1$, which requires $t_\ast|_{\Delta=1}\lneq t_\ast|_{\Delta\ge2}$. 
This holds when
\begin{equation}
\ln\frac{h_{\ast}^{2}}{\T\sqrt{\frac{\pi}{2}}\sqrt{\frac{t_{0}}{\gamma}}}\lneq 4\sqrt{2}\left(1+\tfrac{\kappa_m}{\kappa_d}\right)^2 t_0\gamma.
\end{equation}
Note that this condition is only violated for exponentially small temperatures.

The supercurrent profile $J(t)$ is numerically simulated and plotted in Fig.~\ref{fig:JforV} for various voltages.
To proceed with our analytic analysis, we evaluate the difference from the solution at infinite dissipation $J_\infty(t)$ given by
\begin{align}
&\kappa_m<\kappa_d: && J_{\infty}(t)=J_{0}\begin{cases}
\frac{t}{t_{0}} & t<t_{\mathrm{init}},\\
\frac{t}{t_{0}}-2\frac{t-t_{\mathrm{init}}}{t_{\mathrm{fin}}-t_{\mathrm{init}}} & t_{\mathrm{init}}<t<t_{\mathrm{fin}},\\
\frac{t-2t_{0}}{t_{0}} & t_{\mathrm{fin}}<t,
\end{cases}\nonumber\\
&\kappa_m\geq\kappa_d: && J_{\infty}(t)=J_{0}\begin{cases}
\frac{t}{t_{0}} & t<t_{\mathrm{init}},\\
\frac{t-2t_{0}}{t_{0}} & t>t_{\mathrm{init}}.
\end{cases}
\end{align}
We approximate the unwinding events as immediate transitions to $J_{\infty}$ at time
$t_{\ast}$; this makes our following results hold up to order-one corrections:
\begin{align}\label{eq:Jmodel}
\bar{J} & \sim \frac{1}{2t_{0}}\int_{0}^{t_{\ast}}dt\left\{ J_{0}\frac{t}{t_{0}}-J_{\infty}(t)\right\} \nonumber \\
& = J_{0}\begin{cases}
\frac{\left(t_{\ast}-t_{\mathrm{init}}\right)^{2}}{2\left(t_{\mathrm{fin}}-t_{\mathrm{init}}\right)t_{0}} & t_{\ast}\le t_{\mathrm{fin}},\\
\frac{t_{\ast}-t_{0}}{t_{0}} & t_{\ast}\ge t_{\mathrm{fin}}.
\end{cases}
\end{align}
Plugging Eq.~(\ref{eq:t_ast}) into Eq.~(\ref{eq:Jmodel}) immediately yields
\begin{widetext}
\begin{equation}\label{eq:Jcasesapp}
\bar{J} \sim J_{0}\begin{cases}
\left(1-\frac{\kappa_m}{\kappa_d}\right)^{-1}\frac{1}{4t_{0}\gamma}\ln\frac{h_{\ast}^{2}}{\T\sqrt{\frac{\pi}{2}}\sqrt{\frac{t_{0}}{\gamma}}} & \quad\text{for}~\sqrt{\frac{1}{2t_{0}\gamma}\ln\frac{h_{\ast}^{2}}{\T\sqrt{\frac{\pi}{2}}\sqrt{\frac{t_{0}}{\gamma}}}}<1-\frac{\kappa_m}{\kappa_d},\\
\frac{1}{2}\left(\frac{\kappa_m}{\kappa_d}-1\right)+\sqrt{\frac{1}{2t_{0}\gamma}\ln\frac{h_{\ast}^{2}}{\T\sqrt{\frac{\pi}{2}}\sqrt{\frac{t_{0}}{\gamma}}}} & \quad\text{for}~\sqrt{\frac{1}{2t_{0}\gamma}\ln\frac{h_{\ast}^{2}}{\T\sqrt{\frac{\pi}{2}}\sqrt{\frac{t_{0}}{\gamma}}}}\ge1-\frac{\kappa_m}{\kappa_d}.
\end{cases}
\end{equation}
\end{widetext}
Here, recall that $h_{\ast}\thicksim\frac{\pi}{2}$, and note, that for $\kappa_m=\kappa_d$, the exact solution Eq.~\eqref{eq:J_exact} matches the form above for $h_{\ast}=2\sqrt{2}$.

To evaluate the consistency condition in Eq.~(\ref{eq:forget}), we use $t_{\mathrm{prev}}=\max\{t_\ast,t_{\mathrm{fin}}\}-2t_0$ as well as $\frac{t_{\ast}}{t_{0}}\sim\frac{\bar{J}}{J_{0}}+1$ for $t_{\ast}\ge t_{\mathrm{fin}}.$
We thus find
\begin{equation}
\frac{1}{\sqrt{2\pi t_0\gamma}}\ll\begin{cases}
1+\frac{\kappa_m}{\kappa_d} & t_{\ast}\le t_{\mathrm{fin}},\\
\frac{1}{2}\left(3+\frac{\kappa_m}{\kappa_d}\right)-\frac{\bar{J}}{J_{0}} & t_{\ast}\ge t_{\mathrm{fin}}.
\end{cases}
\end{equation}
This is the condition on moderately low currents presented in Sec.~\ref{sec:trans}.

Our results in Eq.~\eqref{eq:Jcasesapp} may be re-expressed using the original model parameters and $\bar{J}_c=\frac{2\pi q(\kappa_m-\kappa_d)}{L_x}$. This yields the characteristic $I$-$V$ curves of the model in Eq.~(\ref{eq:JV}),
\begin{widetext}
\begin{align}\label{eq:Jcases}
\kappa_m<\kappa_d:\qquad 
&\begin{cases}
\bar{J}\propto q\frac{L_x \kappa_d}{\Gamma(\kappa_d-\kappa_m)}qV &
\textstyle\quad\text{for }\bar{J}<|\bar{J}_c|\text{ and }
\textstyle\frac{\kappa_m}{\kappa_d}+1\gg\frac{L_x}{4\pi^2}\sqrt{\frac{qV}{\Gamma\kappa_d}}, \\
\bar{J}-\bar{J}_c\propto q\sqrt{\frac{\kappa_d}{\Gamma}qV} &\textstyle\quad\text{for }\bar{J}>|\bar{J}_c|\text{ and }
\textstyle\frac{1}{2}(\frac{\kappa_m}{\kappa_d}+3)-\frac{\bar{J}}{J_{0}}\gg\frac{L_x}{4\pi^2}\sqrt{\frac{qV}{\Gamma\kappa_d}},
\end{cases} \nonumber\\
\kappa_m\ge\kappa_d:\qquad 
&\begin{cases}\bar{J}-\bar{J}_c\propto q\sqrt{\tfrac{\kappa_d}{\Gamma}qV} &\textstyle\quad\text{for }\frac{1}{2}(\frac{\kappa_m}{\kappa_d}+3)-\frac{\bar{J}}{J_{0}}\gg\frac{L_x}{4\pi^2}\sqrt{\frac{qV}{\Gamma\kappa_d}}.
\end{cases}
\end{align}
\end{widetext}

%%%%%%%%%%%%%%%%%%%%%%%%%%%%%%%
\begin{figure}[!t]
    \centering
    \includegraphics[width=1\linewidth]{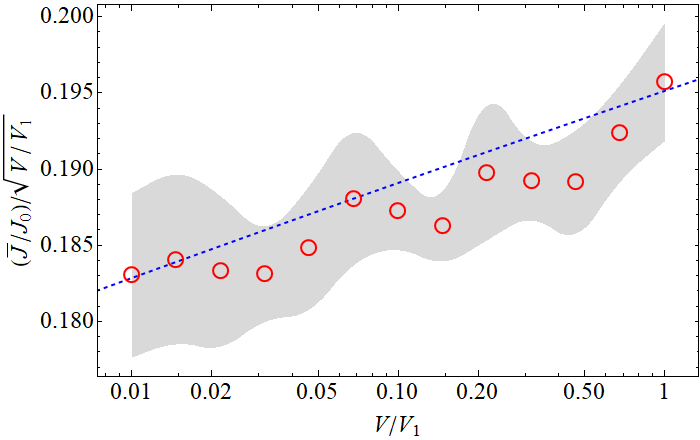}
    \caption{Logarithmic corrections to the current-voltage curve for $\kappa_m=\kappa_d$; see Fig.~\ref{fig:exact-fit}. Each circle depicts the average supercurrent from 18 instances of numerical simulations of Eq.~(\ref{eq:TDGLcross}); the gray area depict 95\% statistical confidence. The dashed line depicts the analytic results of Eq.~(\ref{eq:J_avg_exact}). Here, $V_1=\frac{\kappa_d\Gamma}{qL_x^2}$. There are no fitting parameters.}
    \label{fig:exact-fit-dev}
\end{figure}
%%%%%%%%%%%%%%%%%%%%%%%%%%%%%%%

\section{Numeric solutions to the TDGL equations}\label{app:numeric}

In this Appendix, we provide the details of the numerical simulations presented in Sec.~\ref{sec:sol}.

We numerically evaluate
Eq.~(\ref{eq:transport-PDE}) for a time interval $t\in[0,\tau]$.
We use a stochastic noise term
\begin{align}
\bm{\zeta}(x,t)  = \ &\frac{1}{\sqrt{L_xL_y\tau}}\sum_{k_x=-N_x}^{N_x}\sum_{k_y=-N_y}^{N_y}\sum_{\omega=-N_t}^{N_t} \nonumber\\
&\frac{1}{\sqrt{2}}\left(\mathbf{c}_{\mathbf{k},\omega}^{\phantom{|}}+\mathbf{c}_{-\mathbf{k},-\omega}^{\ast}\right)e^{2\pi i\frac{k_x x}{L_x}}e^{2\pi i\frac{k_y y}{L_y}}e^{2\pi i\frac{\omega t}{\tau}},
\end{align}
where $N_x,N_y,N_t$ are integers. Here the noise Fourier components $\{\mathbf{c}_{\vec{k},\omega}\}$ are random variables satisfying
\begin{align}
\tavg{c_{i,\mathbf{k},\omega}^{\phantom{|}}c_{j,\mathbf{k}',\omega'}^{\ast}}   = \ & \T\delta_{ij}\delta_{\mathbf{k},\mathbf{k}'}\delta_{\omega,\omega'}.
\end{align}
This formulation ensures that
\begin{align}
&\avg{\zeta_i(\rr,t)\zeta_j(\rr,t)}   =\T\delta_{ij}(2N_x+1)(2N_y+1)(2N_t+1),\\
&\int_{0}^{L_x}dx\int_{0}^{L_y}dy\int_{0}^{\tau}dt\avg{\zeta_i(\rr,t)\zeta_j(\rr',t')}   =\T\delta_{ij}.
\end{align}

\subsection{Exact parameter values for the figures}

In Fig.~\ref{fig:exact-fit} we choose $L_y/L_x=1$, $\kappa_m/\kappa_d=1$, $\T/\gamma=\frac{1}{4\pi^2}\times 10^{-7}$, $qV_1/\gamma=\frac{1}{4\pi^2}$, $V/V_1\in[10^{-2},1]$, $\tau=\frac{4\pi}{qV}$, $N_x=2$, $N_y=0$, $ N_t=100$.
We compare the results to the analytic expression of Eq.~(\ref{eq:J_avg_exact});
note that there are no fitting parameters.

In Fig.~\ref{fig:JforV} we choose $L_y/L_x=1$, $\kappa_m/\kappa_d=0$,  $\T/\gamma=\frac{1}{4\pi^2}\times 10^{-9}$, $qV_1/\gamma=\frac{1}{2\pi^2}$, $V/V_1\in[1,4]$, $\tau=\frac{4\pi}{qV_1}$, $N_x=8$, $N_y=0$, $ N_t=100$.

In Fig.~\ref{fig:exact-fit-dev} we normalize Fig.~\ref{fig:exact-fit} by the leading power-law behavior and find a very good agreement with our predicted logarithmic corrections.

\end{document}